\documentclass[10pt,journal,onecolumn,romanappendices]{IEEEtran}
\usepackage{cite,calc}
\usepackage[cmex10]{amsmath}
\usepackage{amsfonts,psfrag,amssymb,graphicx,subfigure}
\usepackage[mathcal]{euscript}  

\newcommand{\thetaD}[1]{\Theta(#1)}

\newcommand{\SNR}{\mathrm{SNR}}

\newcommand{\pr}{\mathbb{P}}
\newcommand{\Pre}{\Prm_\mathrm{e}}
\newcommand{\Preb}{\overline{\Prm}_\mathrm{e}}
\newcommand{\Prm}{P}
\newcommand{\Prmb}{\overline{\Prm}}
\newcommand{\Prmr}{\Prm_\mathrm{region}}
\newcommand{\Prmrb}{\Prmb_\mathrm{region}}
\newcommand{\Prmu}{\Prm_\mathrm{union}}
\newcommand{\Prmub}{\Prmb_\mathrm{union}}
\newcommand{\Rcone}{\cR_\mathrm{c}}
\newcommand{\Rsphere}{\cR_\mathrm{s}}
\newcommand{\asphere}{\alpha_\mathrm{s}}
\newcommand{\bcone}{\beta_\mathrm{c}}
\newcommand{\RML}{\cR_\mathrm{ML}}
\newcommand{\myDef}{\triangleq}

\newcommand{\Et}{\tilde{E}}
\newcommand{\Ebd}{E_\mathrm{d}}
\newcommand{\Ebdt}{\Et_\mathrm{d}}
\newcommand{\expeq}{ \stackrel{.}{=} }  
\newcommand{\exple}{ \stackrel{.}{\le} }
\newcommand{\expge}{ \stackrel{.}{\ge} }

\newcommand{\Rcrit}{R_\mathrm{crit}}

\newcommand{\RR}{{\mathbb R}}

\newcommand{\E}{\mathbb{E}}

\newcommand{\cV}{\mathcal{V}}
\newcommand{\cR}{\mathcal{R}}
\newcommand{\cB}{\mathcal{B}}
\newcommand{\cC}{\mathcal{C}}
\newcommand{\cD}{\mathcal{D}}
\newcommand{\cE}{\mathcal{E}}
\newcommand{\cH}{\mathcal{H}}

\newcommand{\bzero}{\mathbf{0}}

\newcommand{\bb}{\mathbf{b}}

\newcommand{\bm}{\mathbf{c}}

\newcommand{\bx}{\mathbf{x}}
\newcommand{\bc}{\mathbf{c}}
\newcommand{\be}{\mathbf{e}}
\newcommand{\bu}{\mathbf{u}}
\newcommand{\bv}{\mathbf{v}}
\newcommand{\bw}{\mathbf{w}}
\newcommand{\by}{\mathbf{y}}
\newcommand{\bz}{\mathbf{z}}

\newcommand{\bX}{\mathbf{X}}

\newcommand{\bzperp}{\bz_{\by^{\perp}}}

\newtheorem{theorem}{Theorem}
\newtheorem{corollary}{Corollary}

\newtheorem{lemma}{Lemma}
\newtheorem{prop}{Proposition}

\begin{document}


\title{Geometric Relationships Between\\
       Gaussian and Modulo-Lattice Error Exponents}

\author{Charles~H.~Swannack,
Uri~Erez,~\IEEEmembership{Member,~IEEE,}
Gregory~W.~Wornell,~\IEEEmembership{Fellow,~IEEE} \thanks{This work
  was supported in part by the National Science Foundation under Grant
  No.~CCF-0635191.    This work was presented in part at the Allerton
  Conference on Communication, Control, and Computing, Monticello, IL,
  Sep.\ 2005.}
\thanks{U.~Erez is with the Department of
Electrical Engineering - Systems, Tel Aviv University, Ramat Aviv,
69978, Israel (Email: uri@eng.tau.ac.il).} \thanks{C.~H.~Swannack and
  G.~W.~Wornell are with the Department
of Electrical Engineering and Computer Science, Massachusetts
Institute of Technology, Cambridge, MA 02139 (Email:
\{swannack,gww\}@mit.edu).}}
\maketitle

\begin{abstract}
Lattice coding and decoding have been shown to achieve the
capacity of the additive white Gaussian noise (AWGN) channel.   This
was accomplished using a minimum mean-square error scaling and
randomization to transform the AWGN channel into a modulo-lattice
additive noise channel of the same capacity.   It has been further
shown that when operating at rates below capacity but above the
critical rate of the channel, there exists a rate-dependent scaling
such that the associated modulo-lattice channel attains the error
exponent of the AWGN channel.   A geometric explanation for this result
is developed.   In particular, it is shown how the geometry of typical
error events for the modulo-lattice channel coincides with that of a
spherical code for the AWGN channel.
\end{abstract}


\section{Introduction}
\label{sec:intro}

\PARstart{T}{he} capacity of the additive white Gaussian noise (AWGN)
channel was analyzed by Shannon in 1948 in his foundational
work\cite{ShannonMath}.   In 1959 Shannon subsequently studied lower
and upper bounds on the error exponent achieved by codes for this
channel \cite{ShannErrorProb}.  These bounds, while quite tedious to
derive, relied on simple geometric arguments.

An alternative derivation of these results, which uses methods
originally developed for general discrete memoryless channels (DMC),
was later provided by Gallager in 1965 \cite{GallagerSimple}.   This
derivation, while much simpler from an analytic standpoint lacked much
of the geometry that was contained in Shannon's original work.
Further work by Shannon, Gallager and Berlekamp in 1967
\cite{SGB67-I,SGB67-II} provided a tighter upper bound on the
reliability function for low rates, which was recently improved upon
by Ashikhmin et~al.\ \cite{AlexiUpper00}.

The lower and upper bounds coincide for rates greater than the
critical rate $\Rcrit$ of the channel and therefore the error
exponent is known for rates $\Rcrit<R<C$.  These works further
show that with (optimal) maximum likelihood (ML) decoding, the
sphere-packing exponent can be achieved for $R \ge \Rcrit$ by
random spherical ensembles, i.e., by a code whose codewords are
drawn uniformly over the surface of a sphere.

A different line of work aimed at developing structured codes for the
AWGN channel using lattice codes was initiated by de Buda
\cite{deBuda89}.  It was shown in \cite{UriMLANPaper04} that the use of
lattice codes in conjunction with lattice decoding can indeed achieve
capacity on the AWGN channel.  One of the key elements in the
transmission scheme involves transforming the AWGN channel into an
unconstrained modulo-lattice additive noise channel
(as we describe in Section~\ref{sec:lattice_def}), having
(asymptotically in the dimension of the lattice) the same capacity as
the original channel.  For the resulting channel, if one uses a lattice
code $\Lambda_c$ such that $\Lambda \subset \Lambda_c$, then ML
decoding amounts to lattice decoding of $\Lambda_c$.

A second key ingredient in the mod-$\Lambda$
transformation\footnote{The associated modulo-lattice channel is
typically referred to as the mod-$\Lambda$ channel.} is the use of
scaling, i.e., a linear estimator, at the receiver prior to the
application of the modulo operation.   It was observed in
\cite{UriMLANPaper04} that using minimum mean-square error (MMSE)
scaling minimizes the variance of the noise in the resulting
mod-$\Lambda$ channel and results in a channel with capacity
approaching (in the limit of large lattice dimension) the capacity of
the AWGN channel.  Thus, MMSE scaling is a natural choice and indeed is
unique if one aims for capacity \cite{ForneyMMSE03}.

We note that using a lattice code for transmission over the
mod-$\Lambda$ channel does not incur any penalty in terms of error
exponent of the mod-$\Lambda$ channel \cite{UriMLANPaper04}, i.e., the
error probability (as measured by the best known lower bound on the
error exponent) of a good (possibly randomly generated) lattice code
is no greater than that of a random code.\footnote{This property is a
counterpart to the sufficiency of linear codes for achieving the best
known lower bounds on the error exponent of the binary symmetric
channel \cite{BargForney}.}  Thus, it suffices to study the error
exponent of the mod-$\Lambda$ channel, which may be done by standard
random coding arguments.  Indeed, the error exponent of the
mod-$\Lambda$ channel is interesting in its own right, as it plays a
key role in other problems as well.  For instance the error exponent
of the mod-$\Lambda$ channel provides a lower bound on the error
exponent of the dirty-paper channel \cite{CostaDirtyPaper} for
arbitrarily strong interference \cite{UriMLANPaper04}.

It was further conjectured by the authors of \cite{UriMLANPaper04}
that the mod-$\Lambda$ transformation, while not incurring a loss in
mutual information, does incur a loss in error exponent.  Recently,
however, Liu et~al.\ \cite{LiuMLAN} have shown that while MMSE scaling
is not sufficient to obtain the error exponent for the mod-$\Lambda$
channel, a \emph{different} scaling is nonetheless sufficient to
obtain the random coding exponent.  Through some quite rigorous
computation, \cite{LiuMLAN} shows that using a rate-dependent scaling,
the sphere packing (i.e., optimal) error exponent can be achieved for
rates exceeding the critical rate of the AWGN channel.

The goal of the present work is to provide a unifying geometrical
framework for the derivation of the error exponents for both AWGN and
mod-$\Lambda$ channels.  We obtain a simple explanation for the
results of \cite{LiuMLAN}, and in particular to the scaling that
maximizes the error exponent of the mod-$\Lambda$ channel at high
rates.  We use geometric arguments in order to study the typical error
events in both the AWGN and mod-$\Lambda$ channels.  We start by
analyzing random spherical codes and observe that the optimal
mod-$\Lambda$ scaling occurs naturally in this context as well.  We
develop a simple geometrical picture of the relationship between the
typical error events in the mod-$\Lambda$ and AWGN channels via
identification of transmitted codewords.

At low rates the best known bounds for the error exponent are based on
minimum distance arguments.  The identification of transmitted
codewords plays a key role in our development of the error exponent
for the mod-$\Lambda$ in this regime, a region not explicitly
characterized in \cite{LiuMLAN}.  We show that in this region there is
a rate dependent scaling using which the error exponent of an ensemble
of lattice codes matches the error exponent of an ensemble of
spherical codes, provided that both code ensembles have been
expurgated to meet the same minimum distance criterion.  More
precisely, the error exponent for an ensemble of (expurgated)
spherical codes is equal to the error exponent for an ensemble of
lattice codes provided that every code in each of these two ensembles
have the same minimum distance.  However, as the best known bound for
the minimum distance of a spherical code exceeds that of a lattice,
the resulting bound for the error exponent of the mod-$\Lambda$
channel is shown to be less than that in the AWGN channel.  Therefore,
at low rates, the lower bound for the error exponent of the
mod-$\Lambda$ channel is less that that of the AWGN channel.

Another contribution of the present work is the derivation of
exponentially tight bounds for the probability of a mixture of a
spherical noise and AWGN noise leaving a sphere, which we require in
our analysis.  Beyond their use in this work, we believe they may also
be useful in the analysis of other communication problems.  In
particular, it is known that quantization noise arising from a ``good''
high-dimensional quantizer behaves in much the same way as spherical
noise \cite{UriMLANPaper04}.  Thus, the bounds derived in this work
may be useful for the study of error probabilities in communication
systems where both quantization and AWGN noise are present; see, e.g.,
\cite{KochmanWornelISIT10} for a recent such application.

\section{The AWGN Channel: Capacity and Error Exponent} \label{sec:prelim}

In the AWGN channel of interest, the received signal is
\begin{equation}
Y_i=X_i+Z_i,\qquad i=1,2,\dots,n,
\label{AWGNchannel}
\end{equation}
where
\begin{equation*}
\bx = (X_1,X_2,\dots,X_n)
\end{equation*}
is the transmitted signal of length $n$, which satisfies the power
constraint $(1/n) \|\bx\|^2 \leq P$, and where the noise
$Z_1,\dots,Z_n$ are independent, identically distributed (i.i.d.)
Gaussian random variables with zero mean and variance $\sigma^2$.  For
$(n,R)$ codes, i.e., codebooks $\cC$ of $2^{nR}$ codewords, each of
length $n$, the largest rate $R$ such that vanishing error probability
can be achieved is the channel's capacity, which is given by
\begin{equation}
C=\frac{1}{2}\log(1+\SNR),
\end{equation}
where ${\SNR}=P/\sigma^2$ is the signal-to-noise (SNR) ratio on the
channel.

There are many ways to generate random codebook ensembles in
$n$-dimensional space that asymptotically achieve the capacity of the
power constrained AWGN channel.  Possible choices are: an i.i.d.\
Gaussian codebook, a codebook drawn uniformly over the interior of an
$n$-dimensional sphere, a codebook drawn uniformly over the surface of
that sphere, as well as a codebook drawn uniformly over the Voronoi
region of a lattice that is ``good for quantization''
\cite{UriMLANPaper04}.  In essence, the codebook distribution should
approach Gaussianity in a (Shannon) entropy sense, i.e., its entropy
(for a given power) should be close to maximal.

A second-order figure-of-merit for a channel is the error exponent (or
reliability function) of the channel, defined as
\begin{equation*}
   E(R) = \limsup_{n \to \infty} \frac{ - \log {\Pre(n,R)}}{n},
\end{equation*}
where ${\Pre(n,R)}$ is the minimal value of the probability of error
$\Pre(\cC)$ over all $(n,R)$ codes $\cC$, and where, in turn,
$\Pre(\cC)$ is the error probability of a given code $\cC$ averaged
over all codewords.  The error exponent is more sensitive to the
particular choice of codebook input distribution than the channel
capacity.

\begin{figure}[tbp]
   \begin{center}
   \begin{psfrags}
      \psfrag{A}[lc]{sphere-packing}
      \psfrag{B}[lc]{min distance}
      \psfrag{C}[rc]{straight-line}
      \psfrag{D}[bc]{expurgated}
      \psfrag{E}[lc]{mod-$\Lambda$}
      \psfrag{X}[tc]{$R/C$}
      \psfrag{T}[tc]{}
\psfrag{Y}[cc]{$E(R)/\SNR$}
      \psfrag{Rx}[ct]{$R_x$}
      \psfrag{Rc}[ct]{$\Rcrit$}
      \includegraphics[scale=0.5]{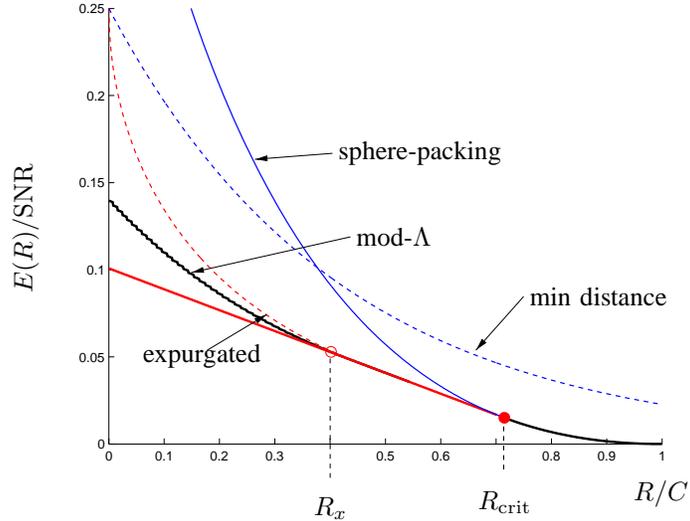}
   \end{psfrags}
   \end{center}
   \caption{Some known bounds on the error exponent for the AWGN
     channel ($\SNR$=$10$ dB).  For rates greater than $\Rcrit$, the
     error exponent for the AWGN channel is the sphere-packing error
     exponent.  For rates less than the critical rate $\Rcrit$, there
     are several known lower bounds, the best known of which is the
     maximum of the random coding error exponent and the expurgated
     error exponent.
     \label{fig:awgn_exp}}
\end{figure}

The error exponent for the AWGN channel is still not known for all
rates.  For rates greater than the critical rate
\begin{equation*}
   \Rcrit = 1/2 \log\left( \frac{1}{2} + \frac{\SNR}{4} +
   \frac{1}{2} \sqrt{1 + \frac{\SNR^2}{4}} \right),
\end{equation*}
the error exponent for the AWGN channel is the sphere-packing error
exponent $E_\mathrm{sp}(R;\SNR)$.  For rates less than $\Rcrit$, there
are several known upper and lower bounds, some of which are depicted in
Fig.~\ref{fig:awgn_exp}.

Geometrically, the sphere-packing error exponent is the exponent of
the probability that the received vector falls outside a cone with
solid angle equal to the average solid angle of an ML decoding region
(i.e., $\exp(-n R )$ times the surface area of the unit sphere).
Recall that the sphere-packing error exponent of the channel
\eqref{AWGNchannel} is \cite{GallagerInfoBook}
\begin{equation}
  E_\mathrm{sp}(R;\SNR) =  E_\mathrm{G}(\beta_G,\rho_G; \SNR),
\label{eqn:sphere_exp}
\end{equation}
where
\begin{align}
 E_\mathrm{G}(\beta,\rho; \SNR)
&\myDef \frac{1}{2} \Bigl[(1-\beta)(1+\rho) + \SNR + \rho \log \beta \notag \\
&\quad\qquad\qquad {}+\log\left(\beta - \frac{\SNR}{1+\rho}\right) - 2 \rho
   R \Bigr],
\label{eqn:gall_Eg}
\end{align}
and with $\beta_G = e^{2 R}$ and
\begin{equation}
      \rho_G = \frac{\SNR}{2 \beta_G} \left(
                       1 +   \sqrt{1 + \frac{4 \beta_G}{\SNR(\beta_G-1)}}
                                          \right) -1.
\label{eqn:rho_g}
\end{equation}

As is well known, a Gaussian codebook does not achieve the error
exponent of the channel, due to the impact of atypical codewords.  As
shown in \cite{ShannErrorProb,GallagerSimple} the error exponent
$E_\mathrm{sp}(R;\SNR)$ is achieved by a codebook drawn uniformly over
the surface of a sphere.\footnote{In \cite{GallagerSimple} Gallager
starts with a Gaussian distribution but applies expurgation to the
same effect.}  We now re-derive the sphere-packing error exponent
using simple geometric arguments in a way that highlights the
relationship between the typical error events in the AWGN and
mod-$\Lambda$ channels.

\section{Geometric Derivation of Sphere-Packing Exponent}
\label{sec:geom_awgn}

In the sequel, we use random coding arguments to bound the probability
of error.  We use $\Omega^{(n)}_0$ to denote the ensemble of codes for
which codewords are drawn independently and uniformly from the surface
of a sphere, and use $\cC$ to denote any codebook drawn from
$\Omega^{(n)}_0$.  We use $\bc$ to denote the transmitted codeword,
and denote any other codeword by $\bc_e$, so that $\cC = \{ \bc \}
\cup \{ \cup_e \, \bc_e \}$.  With $\by=\bc+\bz$ denoting the received
vector, we have than an error occurs under ML decoding when
\begin{equation}
\| \by-\bc_e\| \leq \|\by-\bc\|
\end{equation}
for some codeword $\bc_e \in \cC \setminus
\bc$.  The error probability $\Pre(\bc)$ given that the message $\bc$ is
transmitted is then
\begin{align}
   \Pre(\bc) &=  \pr \left\{
      \| \by  - \bc_e \| \leq \| \by - \bc  \|,
\text{ for some} \  \bc_e \in \cC \setminus \bc  
                      \right\} \notag \\
&=  \pr \left\{ \by  \notin \RML(\bc) \right\}, \notag
\end{align}
where
\begin{equation*}
\RML(\bc)=\{\bx: \|\bx-\bc\| < \|\bx-\bc_e\| \text{ for
  all} \  \bc_e\neq \bc  \}
\end{equation*}
is the ML decoding region of the codeword.

We denote by $\Preb$ the average of $\Pre(\cC)$ over the codebook
ensemble $\Omega^{(n)}_0$.

The sphere-packing (or in the present context ``cone-packing'') lower
bound on the probability of error is straightforward to derive
\cite{ShannErrorProb}.  Indeed let $\Rcone(\bc)$ denote a cone with
apex at the origin and axis passing through $\bc$ such that its volume
is equal to that of $\RML(\bc)$.  Then it easy to show that
\begin{align*}
\Pre(\bc) &=  \pr\left\{ \by  \notin \RML(\bc) \right\} \\
 &\geq \pr\left\{ \by  \notin \Rcone(\bc) \right\}.
\end{align*}
In effect, the ML decoding region cannot be better than a cone with
equal volume.  It further follows by convexity that, for any codebook,
\begin{equation}
\Pre(\cC) \geq  \pr\left\{ \by  \notin \bar{\cR}_c(\bc) \right\},
\label{sp_def}
\end{equation}
where the cone $\bar{\cR}_c(\bc)$ (with apex at the origin and axis
running through $\bc$) has volume equal to the \emph{average} of the
volumes of the ML decoding regions.

The bound \eqref{sp_def} is the well known sphere-packing bound and
when evaluated explicitly (as will be done below) yields the
expression \eqref{eqn:sphere_exp}.  Because the sphere-packing bound
is tight at sufficiently high rates, the ML decoding region may be
well approximated (as far as error probability goes) by a cone.  To
establish the tightness of the bound we next turn to upper bounding
the probability of error.

\subsection{Gallager's  Bounding Technique}

In the following sections we use the method due to Gallager
\cite{GallagerThesis} to bound the probability of decoding error.
Recall that in general the probability of error can be upper bounded
by considering only pairwise errors for all codewords.  More
precisely, the \emph{union bound} yields
\begin{equation}
   \Pre(\bc) \le \sum_{ \bc_\mathrm{e} \ne \bc } \pr\left(
                         \| \by - \bc_e \| \leq
                         \| \by - \bc \|
                      \right) \myDef \Prmu(\bc).
\label{eqn:union_prob_bound}
\end{equation}

While the union bound is tight for low rates (i.e., when the rate is
kept fixed while the SNR and hence the capacity approach infinity) we
require a more general bound in the sequel.  Toward this end, let
$\cR(\bc)$ be a region in $\RR^n$.  Then, more generally, one can
bound the probability of decoding error for a transmitted codeword
$\bc$ considering separately the probability of error when the
received vector is and is not in $\cR(\bc)$.  When $\by\in\cR(\bc)$, we
upper bound the probability of error by using a refined union bound
over all codewords in the codebook, where the noise is bounded to lie
within the region $\cR(\bc)-\bc$.  When $\by\not\in\cR(\bc)$, we upper
bound the probability of error by 1.  Specifically, we have, in general,
\begin{align}
   \Pre(\bc)   &=
         \pr\left( \,
                 \text{error} \, , \, \bc + \bz \in \cR(\bc)
                      \right)  \notag \\
                & \qquad{} + \pr\left( \,
              \text{error} \, , \, \bc + \bz \not\in \cR(\bc)
                      \right) \notag \\
           &\le  \pr\left( \,
                  \text{error} \, , \, \bc + \bz \in \cR(\bc)
                      \right)  \notag \\
          &\qquad{} + \pr\left( \bc + \bz \not \in \cR(\bc) \right)  \notag\\
          &\le  \Prmu^r(\bc) + \Prmr(\bc)
 \label{eqn:gallager_region}
\end{align}
where
\begin{equation}
\Prmr(\bc) \myDef \pr\left(\bc + \bz \not\in \cR(\bc) \right)
\end{equation}
and
\begin{equation}
\Prmu^r(\bc) \myDef
\sum_{ \bc_\mathrm{e} \ne \bc } \pr\left(
                         \| \by - \bc_e \| \leq
                         \| \by - \bc \|
                         , \bc + \bz \in \cR(\bc)
                      \right).
\end{equation}

As shown in the following sections, with a proper choice of the region
$\cR(\bc)$, bound \eqref{eqn:gallager_region} is tight
enough to obtain the sphere-packing error exponent of the AWGN
channel.  In fact, we will see that this is possible by taking
$\cR(\bc)=\bar{\cR}_c(\bc)$, i.e., the same region used to
derive the lower bound on the probability of error.

On the other hand, it is known that the standard union bound
\eqref{eqn:union_prob_bound} is not sufficient to obtain the sphere
packing error exponent.\footnote{Note that the standard union bound
corresponds to taking the region $\cR(\bc)$ to be very large.}
However, as shown in later sections, the standard union bound is tight
enough to obtain the expurgated error exponent bound (the best known
lower bound on the error exponent at low rates).

It is important to emphasize that $\cR(\bc)$ is \emph{not} a decoding
region in general.  The region $\cR(\bc)$ can, in fact, be arbitrary for
the above bound to hold.  However, for a random spherical ensemble
there is no loss in restricting $\cR(\bc)$ to be rotationally
symmetric about the axis that passes through the origin and the
codeword.

For the remainder of this paper we assume that $\cR(\bc)$ is
rotationally symmetric about the axis that passes through the origin
and the codeword and that $\cR(\bc)$ is congruent for all codewords,
and thus, for convenience, use $\cR$ for $\cR(\bc)$.  Thus,
$\Prmr(\bc)$ is the same for all codewords $\bc \in
\cC$ and we simply write this as $\Prmr$.  By averaging
over the code and an ensemble of codes we have
\begin{equation} \label{eqn:gall_bound}
\Preb \leq \Prmub^r(\bc) + \Prmr(\bc)
   =  \Prmub^r + \Prmr
\end{equation}
since by averaging over the ensemble of codes the probability of error
is independent of the codeword.

We now take
the region $\cR$ to be the cone\footnote{This form of
  Gallager's bound corresponds to Poltyrev's tangential sphere bound
  \cite{PoltyrevLinearSpect}.}  of half angle $\theta$ with apex at
the origin and whose axis passes through $\bc$.  We denote this
region as $\Rcone(\theta)$.  Let $\theta_\mathrm{ML}(R)$ be the
angle such that the cone $\Rcone(\theta_\mathrm{ML}(R))$ has a
solid angle equal to the average solid angle of an ML decoding region
for a rate $R$ code.  Thus $\Rcone(\theta_\mathrm{ML}(R))=\Rcone(\bc)$
  as used in the lower bound \eqref{sp_def}.
It is well known \cite{ShannErrorProb} that
\begin{equation*}
\theta_\mathrm{ML}(R) \expeq \theta(R)
\end{equation*}
where $\theta(R)$ satisfies $ \sin \theta(R) = \exp( -R ) $ and
$\expeq$ denotes exponential equality.\footnote{Two functions $f(n)$
and $g(n)$ are said to be exponentially equal if $ \lim_{n \to \infty}
\log\left( {f(n)}/{g(n)} \right) = 0 $ provided the two limits exist.
The notation $\expge$ and $\exple$ are defined analogously.}
Thus, for a given half angle $\theta$ we let $R(\theta) = - \log \sin
\theta$.

With this notation, the sphere-packing lower bound is,
\begin{align}
  \Pre(\cC) &\geq
  \pr\left( \bc + \bz \not\in \Rcone(\theta_\mathrm{ML}(R)) \right) \\
   &\expeq
  \pr\left( \bc + \bz \not\in \Rcone(\theta(R)) \right).
\end{align}
Examining \eqref{eqn:gall_bound}, it is clear that
in order to show that the sphere-packing bound is tight
it is sufficient  to show that for $R \ge \Rcrit$ the
following properties hold
\begin{gather}
\pr\left( \bc + \bz \not\in \Rcone(\theta(R))\right)
\expeq e^{-nE_\mathrm{sp}(R)} \label{eqn:prop_1} \\
\Prmub^r \, \exple \,
e^{-nE_\mathrm{sp}(R)}
\text{ for } R>\Rcrit,  \label{eqn:prop_2}
\end{gather}
with $E_\mathrm{sp}(R)$ as defined in \eqref{eqn:sphere_exp}, and
where the choice of $\Rcone(\theta_\mathrm{ML}(R))$ is left
implicit in \eqref{eqn:prop_2}.

\begin{figure*}
\begin{center}
   \begin{tabular}{cc}
   \begin{psfrags}
      \psfrag{R}[cc]{$\sqrt{n P}$}
      \psfrag{Y}[cc]{}
       \includegraphics[scale=0.65]{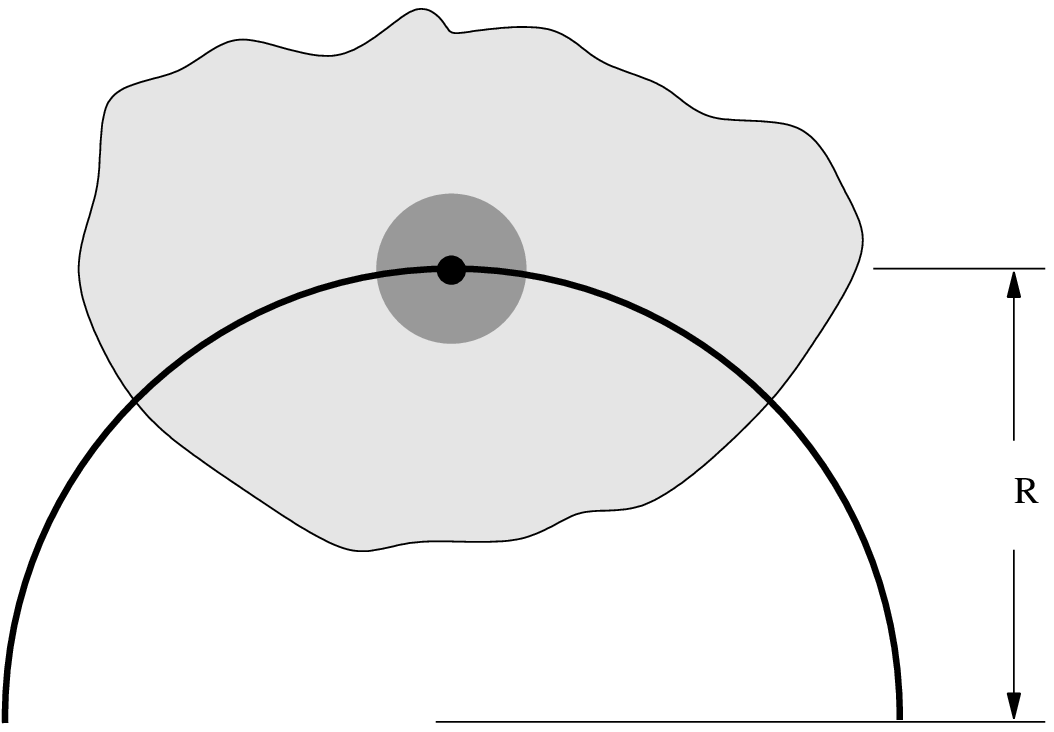}
    \end{psfrags}
     &
   \begin{psfrags}
      \psfrag{T}[lb]{$\theta$}
      \psfrag{R}[cc]{$\cR$}
      \psfrag{B}[cc]{$\beta \sqrt{n P}$}
      \psfrag{Z}[lc]{$y$}
      \psfrag{X}[lc]{${y}^\perp$}
      \psfrag{Y}[cc]{\footnotesize $r(\beta) \sqrt{n P}$}
      \psfrag{z}[lb]{$\bz$}
      \includegraphics[scale=0.65]{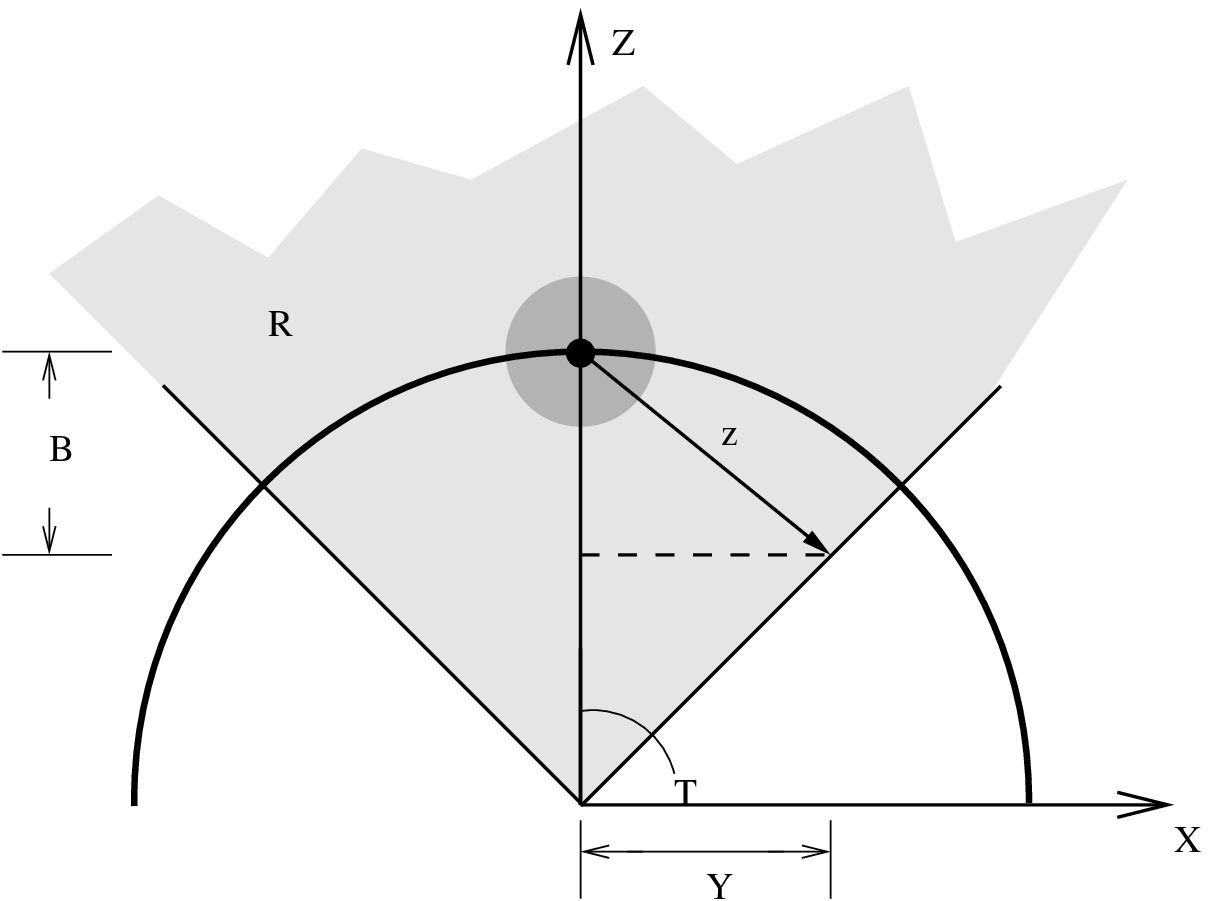}
    \end{psfrags}
   \\
       &
   \\
   (a) & (b)
   \end{tabular}
\end{center}
\caption{A depiction of the general bounding technique
\eqref{eqn:gallager_region}.
  (a) The general bound for an arbitrary region $\cR$ and
  (b) the bounds when specialized to a cone of half angle $\theta$
 \label{fig:gallager_region}}
\end{figure*}

We begin by examining $\pr\left( \bc + \bz \not\in \Rcone(\theta(R))
\right)$ to derive \eqref{eqn:prop_1}.  To do this we follow Berlekamp
\cite{BerlekampErrorCorrect} and decompose the noise into a radial
component $z_y$ normal to the surface of the sphere at $\bc$, and its
orthogonal complement $\bzperp$, as depicted in
Fig.~\ref{fig:gallager_region}.  Then
\begin{equation*}
\bz=z_y \cdot \be_y+\bzperp
\end{equation*}
where $\be_y$ is the unit-vector normal to the sphere at $\bc$.   Let
$\beta \sqrt{nP}$ be the radial component of the noise $z_y$, i.e.,
$\beta = z_y/\sqrt{n P}$, and let $r(\beta) \sqrt{nP}$ be the radius
of the corresponding spherical cross section of the cone $\Rcone(\theta)$, as shown in Fig.~\ref{fig:gallager_region}.  By
simple geometry, we have $r(\beta) = (1+\beta) \tan \theta $.   Thus,
since the components of the noise are independent, we may condition on
the radial component $z_y$ and integrate over the distribution of that
component, i.e.,
\begin{equation}
\label{eqn:int_bound}
   \begin{aligned}
   \pr(\bc &+ \bz \not\in \Rcone )
   = \pr\left( z_y < - \sqrt{n P}  \right) \\
   & \quad + \int_{-1}^{\infty}
p_{z_y}(\beta \sqrt{nP})  \cdot \pr\left( \| \bzperp \| \geq r(\beta) \sqrt{nP}
      \right) d\beta.
   \end{aligned}
\end{equation}

We now apply the following bounds on the norm of a Gaussian vector.
\begin{prop}
\cite{PoltyrevUncon} Let $\bz=(z_1, z_2, \dots, z_n)$ be i.i.d.\
Gaussian random variables with zero mean and variance $\sigma^2$.
Then
\begin{equation*}
   \pr\left( \| \bz \| \ge r \sqrt{n P} \right)
      \expeq
      \begin{array}{lcl}
         \exp\left( -n \, E_h( r^2 P/\sigma^2  ) \right),
      \end{array}
\end{equation*}
where
\begin{equation*}
    E_h( \mu ) \myDef \left\{
     \begin{array}{ll}
        \frac{1}{2}( \mu - 1 - \log \mu  )
      & \text{ if $\mu \ge 1$ } \\
      0 & \text{ otherwise }
     \end{array}
   \right.
\end{equation*}
\end{prop}
In the sequel we examine the exponential behavior of
\eqref{eqn:int_bound}.  First, recall that if $z_1$ is a zero-mean
Gaussian random variable with variance $\sigma^2$, then
\begin{equation*}
   p_{z_1}(\beta \sqrt{nP}) = K_{\sigma}
   \exp\left( -n \, E_v( \beta^2 P/\sigma^2  ) \right)
\end{equation*}
where
\begin{equation*}
      \quad E_v(\mu) \myDef \mu/2
\end{equation*}
and $ K_{\sigma}$ is the normalizing constant.  Now, bounding the
second term on the right-hand side of \eqref{eqn:int_bound} by the
largest term of the integral (as it can be shown the first term on the
right-hand side of \eqref{eqn:int_bound} is never the dominating term)
we have that the probability that the received vector is outside the
cone satisfies
\begin{align}
   &\pr\left( \bc + \bz \not\in \Rcone \right) \notag \\
   &\quad\exple \exp\left( -n \min_{ \beta } \left[
                   E_v( \beta^2 \SNR ) + E_h( r(\beta)^2 \SNR )
                \right] \right).
   \label{eqn:awgn_geom}
\end{align}
It is not hard to show that this inequality is in fact
exponentially tight \cite{ShannErrorProb}.
One can show that in the case that for $0 \le R(\theta) \le C$,
the $\beta$ that minimizes the exponent of \eqref{eqn:awgn_geom} is
\begin{equation} \label{eqn:opt_beta_sphere}
   \beta^*(\theta;\SNR) = \frac{\cos^2 \theta}{2} + \frac{\cos^2 \theta}{2}
      \sqrt{ 1 + \frac{4}{\SNR \cos^2 \theta } } - 1.
\end{equation}
Note that by letting $\theta = \theta(R)$ and
with some additional algebra shown in Appendix \ref{append:rel}
we see that indeed
\begin{equation} \label{eqn:rel}
   E_{v}((\beta^*)^2  \SNR )
                      + E_{h}( r(\beta^*)^2 \, \SNR ) =
   E_\mathrm{sp}(R;\SNR)
\end{equation}
for $R > \Rcrit$.   Thus, \[\pr( \bc + \bz \not\in
\Rcone(\theta(R)) ) \expeq \exp( -n E_\mathrm{sp}(R;\SNR)), \] so that
\eqref{eqn:prop_1} holds.

We verify that \eqref{eqn:prop_2} holds in Section~\ref{sec:low_rates}
and thus for $R> \Rcrit$ the sphere-packing error exponent is
the proper exponent.  To verify \eqref{eqn:prop_2} we employ similar geometric
arguments to those used in this section.  In particular we examine the
typical error events for $\Prmub^r$.  We show
that for all rates greater than the critical rate the error event is
dominated by the same event of leaving a cone at a height of
$\beta^*(\theta(R);\SNR)\sqrt{nP}$.

Note that the probability of leaving the cone at any other height
is in general \emph{smaller} than at $\beta^*$.  Thus, there is some
slack in the choice of the region $\Rcone$.  This leads to
the question of what other regions one can use in order to derive
the sphere-packing bound using Gallager's bounding technique
\eqref{eqn:gallager_region}.

\subsection{Valid Regions for Geometric Derivation}

The tightness of the bound we obtain using the cone
$\Rcone(\theta(R))$ means that for rates greater than the critical
rate, the error probability of leaving an ML decoding region is
exponentially the same as that of leaving the cone
$\Rcone(\theta(R))$.  Thus, the cone approximates\footnote{It is
important to note that the tightness of the sphere-packing bound does
not imply the existence of good cone packings, and in fact these are
known not to exist \cite{kabatyansky78}.}  the ML decoding region in the error
probability analysis for $R > \Rcrit$.  We now investigate how much
freedom we have in choosing a region that has this property.

We refer to any region $\cR$ that yields the sphere-packing bound
as ``valid''.  More specifically, we refer as such to any region for
which the probability that the received vector falls outside this
region is exponentially equal to the sphere-packing error exponent and
is no larger than the cone $\Rcone(\theta(R))$.   More precisely, any
region satisfying the following is valid:
\begin{equation}
\begin{aligned}
&A.  \quad \pr\left(\bc + \bz \notin \cR \right) \expeq
\pr\left(\bc + \bz \notin \Rcone(\theta(R)) \right) \\
&B.  \quad \cR \subset \Rcone(\theta(R))
\end{aligned}
\label{eqn:valid_sphere_def}
\end{equation}
Condition A guarantees that $\Prmr$ remains
exponentially the same for $\cR$ as it was for
$\Rcone(\theta(R))$.  That is, Condition A implies
\eqref{eqn:prop_1}.   Condition B ensures that
$\Prmub^r$ is no greater for $\cR$ than
for $\Rcone(\theta(R))$ and thus if $\Rcone(\theta(R))$ satisfies
\eqref{eqn:prop_2} then by Condition B so does $\cR$.

We seek the ``smallest'' valid region.  Since, as we noted, the cross
section of the cone that dominates the error event is that
corresponding to $\beta^*$, for all other $\beta$ we should be able to
choose a ``narrower'' cross section.  Nonetheless, at a height
$\beta^*$ the radius of $\cR$ has to coincide with that of
$\Rcone(\theta(R))$ since we cannot hope to improve on the
sphere-packing bound.  Thus, $\cR$ must be tangent to
$\Rcone(\theta(R))$ at $\beta^*$.  Further, in order to ensure that
$\pr\left(\bc + \bz \notin \cR \right) \expeq \pr\left(\bc + \bz
\notin \Rcone(\theta(R)) \right)$, it is necessary that a valid region
satisfy, for any $\beta$,
\begin{align}
 &p_{z_y}(\beta) \pr\left( \bc + \bz \not\in \cR \, | \,
      z_y = \beta \sqrt{n P} \right) \notag \\
 &\quad \quad {\leq}
 \, p_{z_y}(\beta^*) \pr\left( \bc + \bz \not\in \Rcone \, | \,
      z_y = \beta^* \sqrt{n P} \right).
   \label{eqn:smallest}
\end{align}

\begin{figure*}
\begin{center}
   \begin{tabular}{cc}
   \begin{psfrags}
      \psfrag{T}[lb]{$\theta(R)$}
      \psfrag{R}[cc]{$\cR$}
      \psfrag{P}[cc]{$\sqrt{n P}$}
      \psfrag{B}[cc]{\footnotesize $\sqrt{n P} \beta^*$}
      \psfrag{Z}[lc]{${z}_1$}
      \psfrag{E}[cc]{\Huge $\star$}
         \includegraphics[scale=0.6]{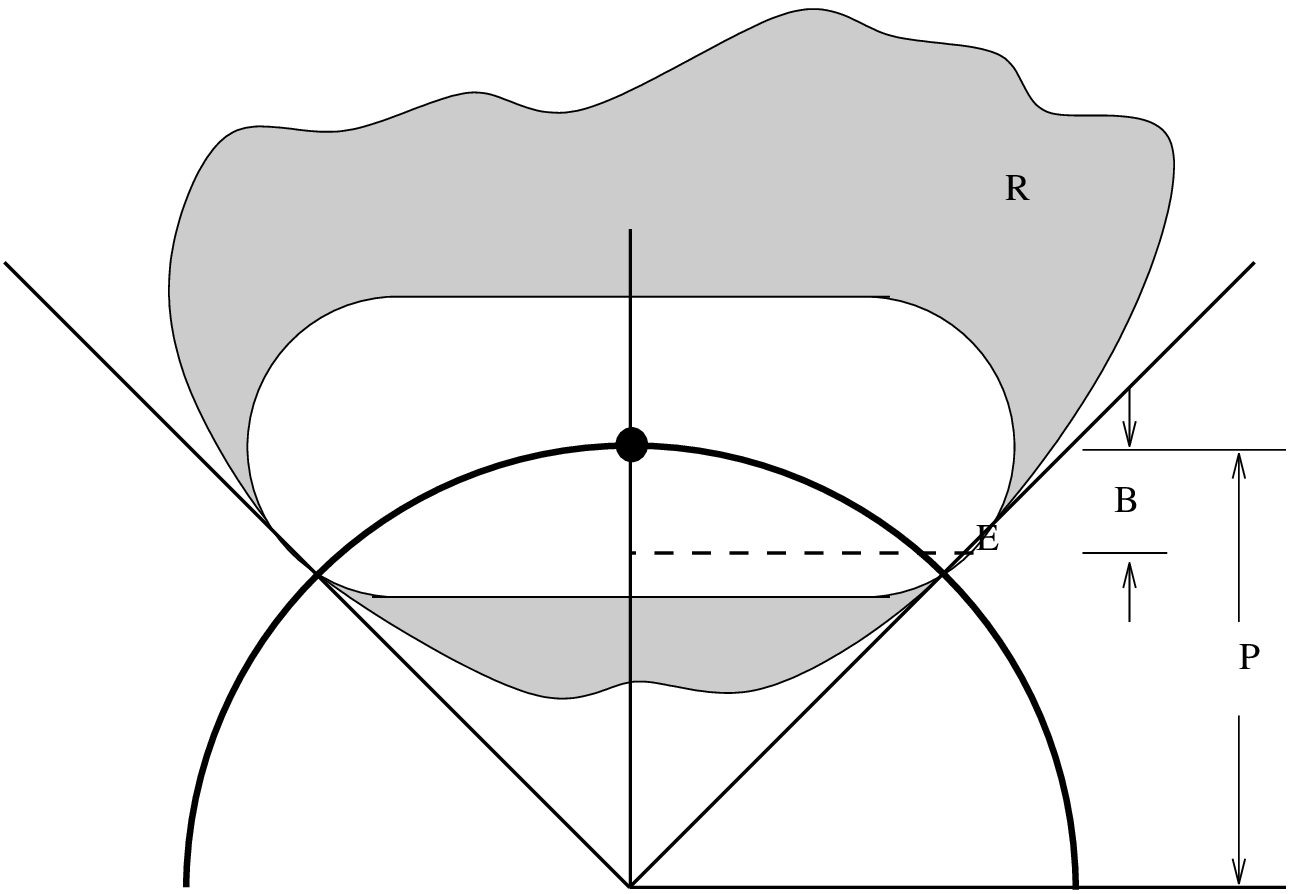}
    \end{psfrags}
     &
   \begin{psfrags}
      \psfrag{T}[lb]{$\theta(R)$}
      \psfrag{R}[cc]{$\Rsphere(\theta(R))$}
      \psfrag{D}[lc]{\footnotesize $d=\frac{\sqrt{n P}}{\asphere^*(R)} e^{-R}$}
      \psfrag{C}[rc]{$\frac\bc{\asphere^*(R)}$}
      \psfrag{B}[cc]{}
      \psfrag{Z}[lc]{${z}_1$}
      \psfrag{E}[cc]{\Huge $\star$}
         \includegraphics[scale=0.6]{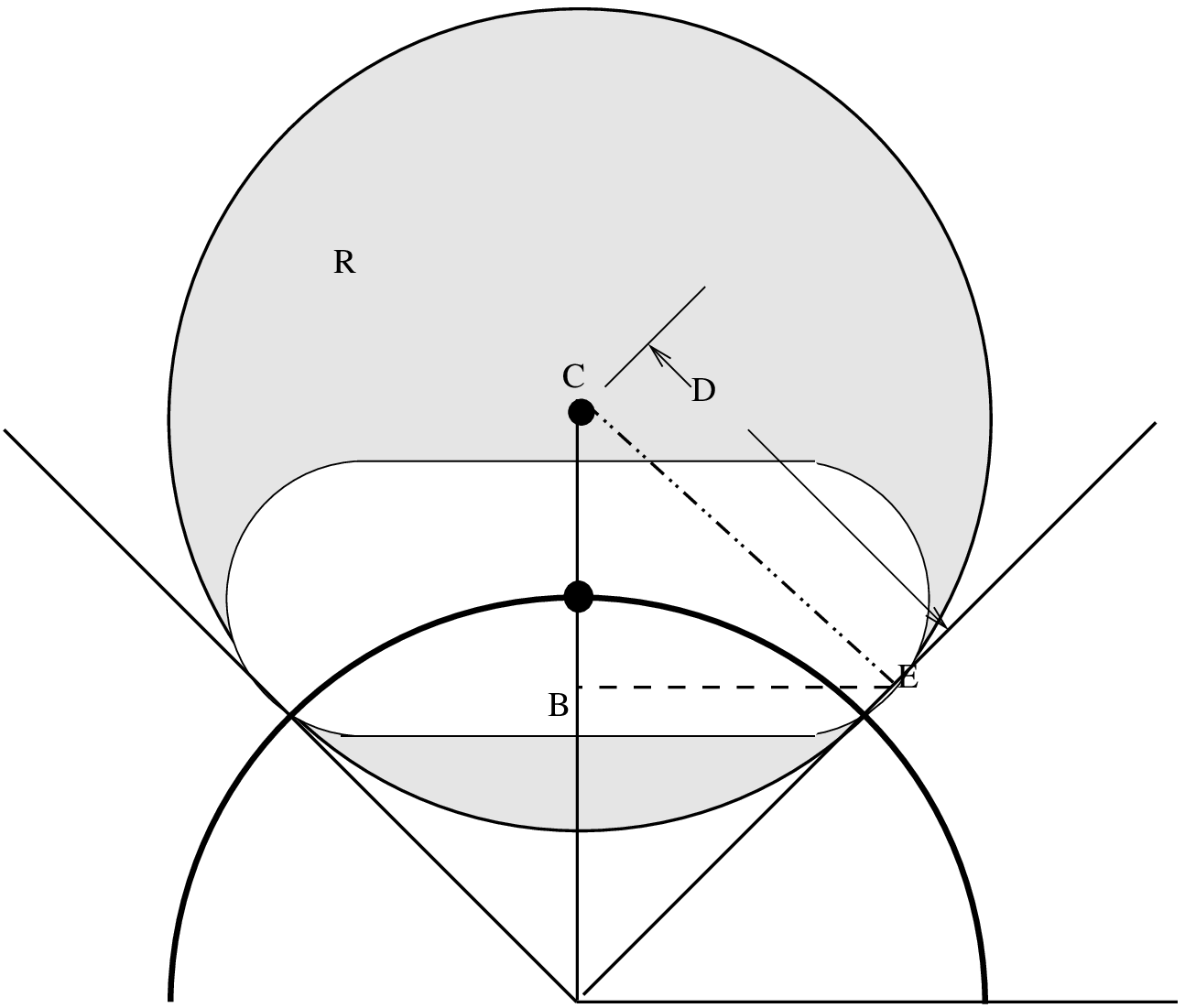}
    \end{psfrags}
   \\
   (a) & (b)
   \end{tabular}
\end{center}
\caption{A depiction of the ``valid'' regions for the general bounding
  technique \eqref{eqn:gallager_region}.  The typical error event
  is depicted by {\Large $\star$}.
  (a) The general bound for an arbitrary ``valid'' region.
      A general region $\cR$ and
       a depiction of the smallest valid region.
  (b) the bounds when specialized to a sphere that contains the smallest
          region.
}
\label{fig:regions}
\end{figure*}

It follows that the region that exactly meets \eqref{eqn:smallest}
with equality for every $\beta$ is the smallest valid region, since
the probability of leaving any other valid region is exponentially
larger.  This region is parametrized by
\begin{equation*}
   E_{v}(\beta^2 \SNR) + E_h( r(\beta)^2 \; \SNR ) =  E_\mathrm{sp}(R; \SNR).
\end{equation*}
As Fig.~\ref{fig:regions}(a) depicts, the smallest region is contained
in a general region $\cR$; all are tangent to the cone at $\beta^*$.

\subsection{Geometric Derivation Using Spherical Regions}
\label{sec:sphere_region}

We now consider the possibility of taking $\cR$ to be a sphere.  This
provides the link to the mod-$\Lambda$ channel.  From the previous
section we know that a valid sphere must be tangent to the cone of
half angle $\theta$ at $\beta^*(\theta;\SNR)$ and must also contain
the smallest region.  As depicted in Fig.~\ref{fig:regions}, in order
to make the sphere tangent at $\beta^*(\theta;\SNR)$ we must draw a
line perpendicular to the cone at $\beta^*(\theta;\SNR)$ and find the
point where this line intersects the line passing through the origin
and the transmitted codeword.  We denote this point (the scaled
codeword) as $\bc/\asphere^*(\theta)$.  Using basic trigonometry we
find that
\begin{equation} \label{eqn:alpha_star}
   \asphere^*(\theta) = \asphere(\beta^*(\theta;\SNR), \theta),
\end{equation}
where
\begin{equation} \label{eqn:alpha_rel}
   \asphere(\beta, \theta) = \frac{ \cos^2 \theta }{1 + \beta}.
\end{equation}
Thus, the radius of this sphere is $\sqrt{n P}/\asphere^*(\theta) \sin
\theta$ and\footnote{Note that this is simply a scaled version of
  $1+\rho_G$.}
\begin{equation} \label{eqn:alpha_star_sphere}
   \frac{1}{\asphere^*(\theta)} = \frac{ 1 + \beta^*(\theta;\SNR)
                        }{ \cos^2 \theta}
                      = \frac{1}{2} \left( 1 + \sqrt{1 + \frac{4}{ \SNR
                           \cos^2 \theta }} \right),
\end{equation}
i.e.,
\begin{equation*}
   \Rsphere(\theta) =  \frac{\sqrt{n P}}{\asphere^*(\theta)} \cdot \be_y
  + \cB\left(\frac{\sqrt{n P}}{\asphere^*(\theta)} \sin \theta\right),
\end{equation*}
where $\cB(r)$ is the ball of radius $r$ and where $\be_y$ is a vector
of unit norm.  We note that $\asphere^*(\theta(R))$ is the optimal
scaling found in \cite{LiuMLAN} for the mod-$\Lambda$ channel.  In
fact, we show in the sequel that the region $\Rsphere(\theta(R))$ is
the natural counterpart of the cone when using Gallager's bounding
technique in the case of the mod-$\Lambda$ channel.

\begin{figure*}
   \begin{center}
      \begin{psfrags}
         \psfrag{X}[rc]{$\bX \in \cC$}
         \psfrag{Y}[bc]{}
         \psfrag{FY}[bc]{}
         \psfrag{F}[cc]{$f$}
         \psfrag{CA}[cc]{}
         \psfrag{Z}[lc]{$\by$ }
         \psfrag{N}[cb]{$\bz$ }
         \psfrag{ML}[cc]{mod-$\Lambda$ }
         \psfrag{QL}[cc]{$Q_{\Lambda/\alpha}(\cdot)$ }
         \psfrag{U}[rc]{$\bb$ }
         \psfrag{P}[cc]{$+$ }
         \psfrag{M}[cc]{$-$ }
         \psfrag{T}[cc]{$\times$ }
         \psfrag{A}[lc]{${1}/{\alpha}$ }
         \psfrag{A2}[cc]{${1}/{\alpha}$ }
 \includegraphics[scale=0.5]{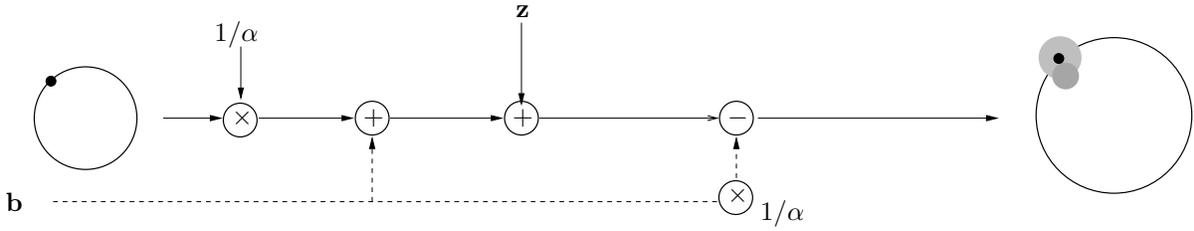}
      \end{psfrags}
   \end{center}
   \caption{A depiction of Lemma~\ref{lem:key_lem}.   A codeword is
     chosen at random from the codebook $\cC$ and scaled by
     $1/\alpha$.  A random dither $ \left( 1 - \frac{1}{\alpha}
     \right)\bb$ is added and the result is transmitted through
     the additive noise channel. \label{fig:lemma}}
\end{figure*}

Before further explaining the connection to the mod-$\Lambda$ channel
we give a second interpretation of our results thus far.   We first
rewrite the received vector as
\begin{align}
\label{eqn:norm_scheme}
      \by  &=  \bc + \bz \\
               &=  \frac\bc{\alpha} +
           \left( 1 - \frac{1}{\alpha} \right) \bc + \bz \\
              &= \frac\bc{\alpha} + \bw
\end{align}
where $\bw$ is a Gaussian vector with mean $(1-1/\alpha) \bc$.  Thus,
we can think of transmission as follows: $\bc/\alpha$ is the chosen
codeword, a deterministic vector of magnitude $1-1/\alpha$ is added to
enable us to meet the power constraint and then the result is
transmitted through the channel where Gaussian noise is added.  Thus,
the probability of leaving the sphere satisfies
\begin{equation*}
\pr\left( \bc+\bz  \notin \Rsphere(\theta) \right)
= \pr\left( \frac\bc{\alpha}+\bw \notin \Rsphere(\theta) \right).
\end{equation*}
Next, let $\bb$ denote a random vector that is uniform over the
surface of a sphere of radius $\sqrt{n P}$ and define the
\emph{effective noise} as
\begin{equation}
\label{eqn:zeff_def}
         \bz_\mathrm{eff} = \frac{1-\asphere^*(\theta)}{\asphere^*(\theta)}\bb +
                \bz.
\end{equation}
From spherical symmetry it now follows that
\begin{equation}
\pr\left( \frac\bc{\alpha}+\bw \notin \Rsphere(\theta) \right) =
\pr\left(\frac\bc{\alpha}+\bz_\mathrm{eff} \notin \Rsphere(\theta)
\right).  \label{effec_equiv}
\end{equation}
Thus, the effect of the deterministic vector
$\left(1-\frac{1}{\alpha}\right)\bc$ is equivalent to that of a
random \emph{spherical noise}.  This ``noise'' is the counterpart of the
``self-noise'' arising in the mod-$\Lambda$ channel as recounted below.
Therefore, the probability of leaving the cone when transmitting the codeword $\bc$ through the channel
(\ref{eqn:norm_scheme}) is the same as when transmitting it through the channel
\begin{align}
\label{eqn:modified_scheme}
            \by_{\rm equiv}     &=  \frac\bc{\alpha} +
           \left( 1 - \frac{1}{\alpha} \right) \bb + \bz.
\end{align}
The channel (\ref{eqn:modified_scheme}) is depicted in Figure~\ref{fig:lemma}.

This leads to the following lemma, establishing that
the sphere-packing error exponent is exponentially equal to the
probability that a random variable that is uniform over the surface of
the ball of radius $(1-\asphere^*(\theta(R)))/{\asphere^*(\theta(R))}
\sqrt{n P}$ plus a Gaussian vector with independent identically
distributed components of variance of $P/\SNR$ remains in a sphere of
radius $\sqrt{n P}/\asphere^*(\theta(R)) \exp(-R)$ about the scaled
codeword.

\begin{lemma}
\label{lem:key_lem}
For $\bz_\mathrm{eff}$ as defined in \eqref{eqn:zeff_def} with
$\asphere^*$ the optimum scaling for the mod-$\Lambda$ channel, and
$\theta(R)$ the half-angle of the cone to which the sphere is
tangent, we have
      \begin{equation} \label{eqn:exp_ball}
      \exp( -n E_\mathrm{sp}(R;\SNR) ) \expeq
         \pr\left(  \bz_\mathrm{eff}
             \not\in \cB\left( \frac{\sqrt{n P}}{\asphere^*(\theta(R))}
                    \sin \theta(R)\right)
         \right)
      \end{equation}
\end{lemma}

\begin{IEEEproof}
This is simple to see using the following exponential equalities
   \begin{equation}
      \begin{aligned}
      & \exp( -n E_\mathrm{sp}(R;\SNR) ) \\
        &\quad \expeq \pr\left(
          \left(1 - \frac{1}{\asphere^*(\theta(R))} \right) \bc + \bz
          \not\in \cB\left( \frac{\sqrt{n P}}{\asphere^*(\theta(R))}
          \sin \theta(R)\right) \right) \\
       &\quad {=}
      \pr\left(
         \left[ \frac{1-\asphere^*(\theta(R))}{\asphere^*(\theta(R))}\bb +
                \bz
         \right]
             \not\in \cB\left( \frac{\sqrt{n P}}{\asphere^*(\theta(R))}
                    \sin \theta(R)\right)
         \right)
      \end{aligned}
      \label{eqn:key}
   \end{equation}
   where the last equality follows from \eqref{effec_equiv}.
   \end{IEEEproof}

\begin{figure*}
  \begin{tabular}{cc}
   \begin{psfrags}
      \psfrag{G}[br]{$r$}
      \psfrag{Y}[bc]{\footnotesize $\sqrt{n P} r(\beta)$}
      \psfrag{B}[cc]{}
      \psfrag{R}[lc]{$\Rsphere(\theta(R))$}
      \includegraphics[scale=0.55]{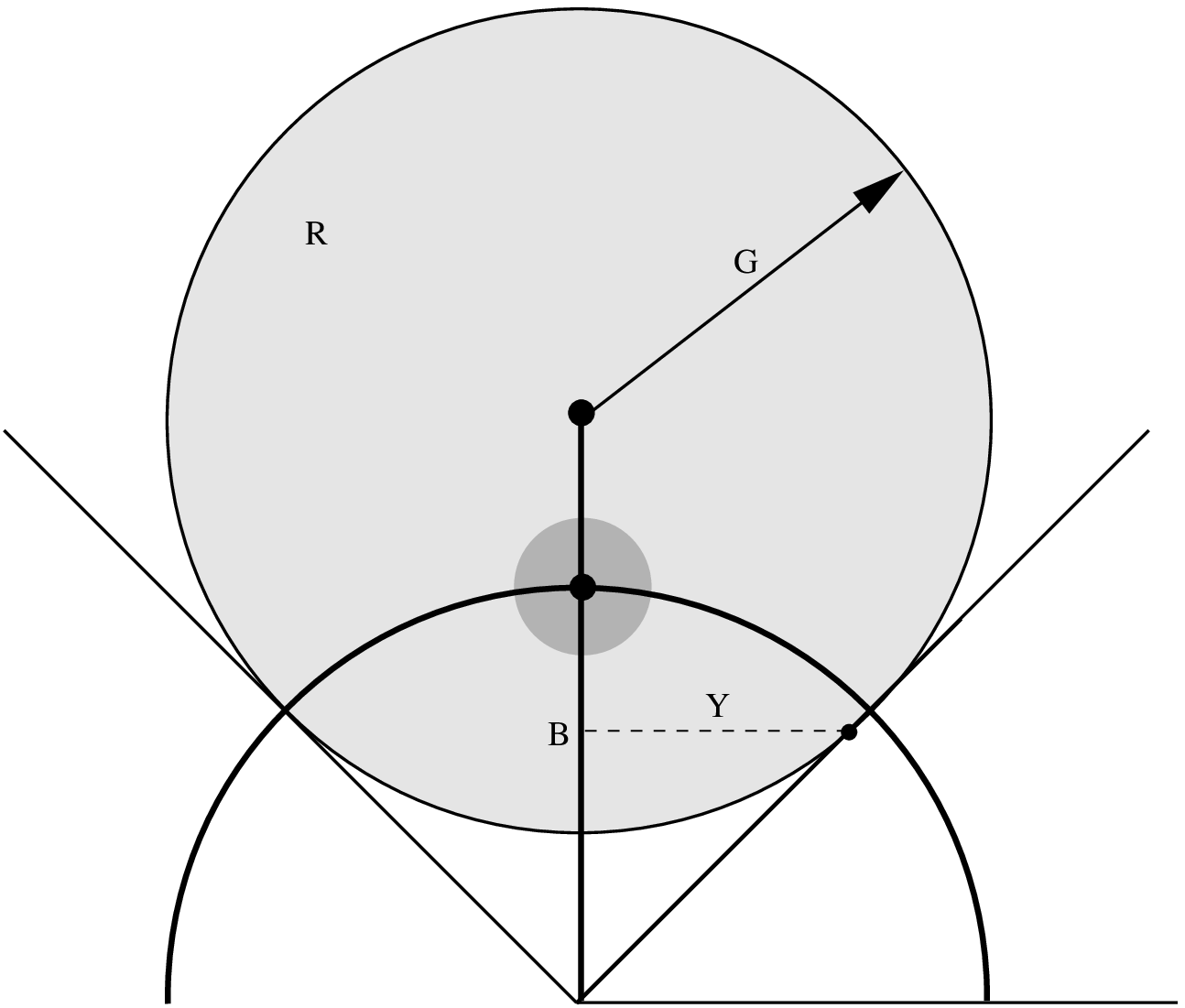}
   \end{psfrags}
&
   \begin{psfrags}
      \psfrag{A1}[cc]{$\frac{1}{\alpha}$}
      \psfrag{A2}[lc]{$1-\frac{1}{\alpha}$}
  \includegraphics[scale=0.55]{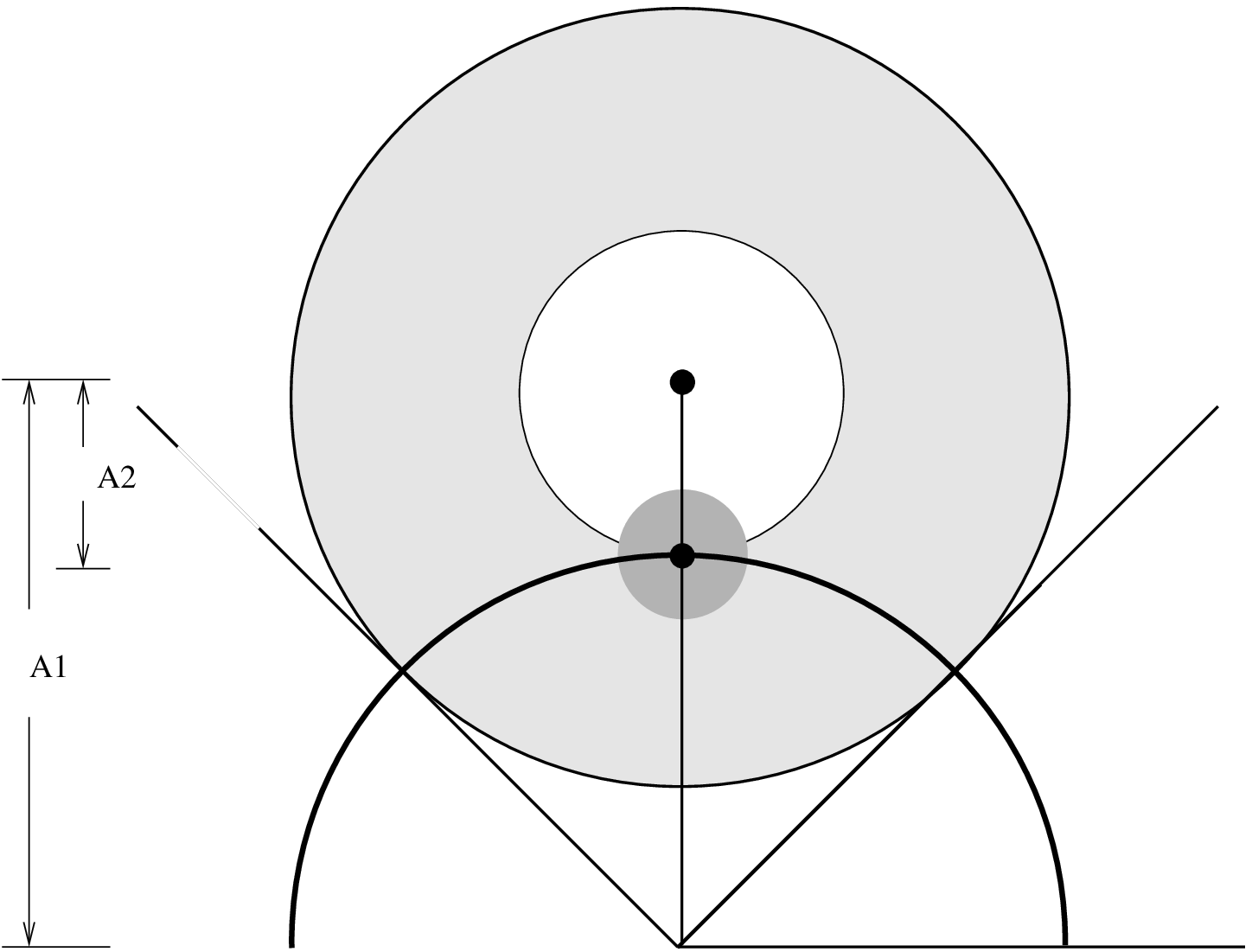}
    \end{psfrags} \\
        (a) & \; \; \; (b)
  \end{tabular}
  \caption{The derivation of the sphere-packing error exponent using a
    spherical region.  (a) The probability of a Gaussian leaving a
    sphere whose center is located at $1/\alpha$ and tangent to a cone
    of half angle $\theta(R)$.  (b) The equivalence between this and
    the probability that a spherical noise plus a Gaussian leaves a
    larger sphere}
    \label{fig:sphere_gauss}
\end{figure*}

The equivalence of Lemma \ref{lem:key_lem} is depicted in
Fig.~\ref{fig:sphere_gauss}.   We make the final connection to the
error probability in the mod-$\Lambda$ channel after briefly
summarizing those aspects of the channel we'll need.

\section{Modulo Lattice Additive Noise Channel} \label{sec:lattice_def}

In \cite{UriMLANPaper04}, a lattice-based transmission scheme was
proposed for the power-constrained AWGN channel.  The scheme
transforms the AWGN channel into a mod-$\Lambda$ channel.  In this
section, we relate the latter to the geometrical derivation of the
AWGN error exponent we developed earlier.  We briefly review the
lattice transmission approach proposed in \cite{UriMLANPaper04}.  We
first recall a few definitions pertaining to lattices.

An $n$-dimensional lattice $\Lambda$ is a discrete subgroup of the
Euclidean space $\RR^n$.   The fundamental Voronoi region of a lattice
$\cV=\cV(\Lambda)$ can be taken as any set such that the following is
satisfied:
\begin{itemize}
\item If $\bx \in \cV$ then $\|\bx-\bzero\| \leq \|\bx-\lambda \|$
for any $\lambda \in \Lambda \setminus \{ 0 \}$.
\item Any point $\by \in \RR^n$ can be uniquely written as
$\by=\lambda+\bx$, where $\lambda \in \Lambda$ and $\bx \in \cV$.
Thus, $\bx$ is the remainder when reducing $\by$ modulo $\Lambda$.
\end{itemize}
Clearly all fundamental regions have the same
volume.   Thus, we let $V = V(\Lambda)$ be the volume of any (every)
fundamental region.  To each lattice we may associate an ``effective radius,''
$r_\Lambda^\mathrm{eff}$, which is the radius of the sphere having the same
volume of $\cV$, i.e.,
\begin{equation}
\label{eqn:reff_def}
   r_\Lambda^\mathrm{eff} =
    \left( \frac{V(\Lambda)}{\mathrm{Vol}(
                   \cB(1))} \right)^{1/n}.
\end{equation}
We next recall two important figures of
merit for any lattice that are required in the sequel; see, e.g.,
\cite{ErezGoodLattice} for a further discussion of these figures
of merit.
\begin{figure*}
   \begin{center}
      \begin{psfrags}
         \psfrag{X}[rc]{$\bX \in \cC$}
         \psfrag{Y}[bc]{} 
         \psfrag{FY}[bc]{}
         \psfrag{F}[cc]{$f$}
         \psfrag{Z}[lc]{$\by$ }
         \psfrag{N}[cb]{$\bz$ }
         \psfrag{ML}[cc]{mod-$\Lambda$ }
         \psfrag{QL}[cc]{$Q_{\Lambda/\alpha}(\cdot)$ }
         \psfrag{U}[rc]{$\bu$ }
         \psfrag{P}[cc]{$+$ }
         \psfrag{M}[cc]{$-$ }
         \psfrag{T}[cc]{$\times$ }
         \psfrag{A}[lc]{${1}/{\alpha}$ }
 \includegraphics[scale=0.5]{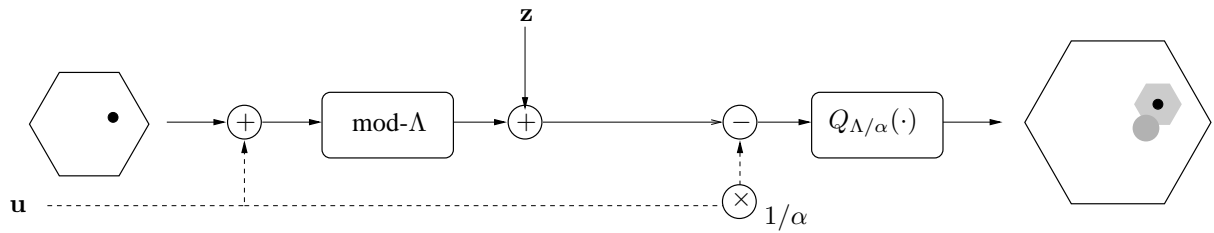}
      \end{psfrags}
   \end{center}
   \caption{The mod-$\Lambda$ channel diagram.  A codeword is chosen
      at random from the codebook $\cC$ and a random dither $\bu$ is
      added and the result (mod-$\Lambda$) is transmitted through the
      additive noise channel.  An estimate of the received signal is
      then formed, the dither subtracted, and the result quantized
      (decoded) to the nearest codeword. \label{fig:mlan}}
\end{figure*}

The normalized second-moment of a lattice $G(\Lambda)$ is
\begin{equation*}
G(\Lambda) \myDef  \frac {\sigma^2(\Lambda)} {|V|^{2/n}},
\end{equation*}
where, in turn, $\sigma^2(\Lambda)$ is the second-moment of the lattice
\begin{equation*}
\sigma^2(\Lambda)
\myDef \frac{1}{n} \frac {\int_{\cV}\|\bx\|^2 d \bx} {|V|}.
\end{equation*}
It is known that $G(\Lambda)$ is always greater than $1/(2 \pi e)$, the
normalized second-moment of a sphere.
Lattices such that $G(\Lambda) \approx 1/(2 \pi e)$
are useful in quantization theory and are said to be ``good for quantization.''
A second important figure of merit of any lattice is its covering radius,
$r_\Lambda^\mathrm{cov}$.  To be precise, recall that the set
$\Lambda + \cB(r)$ is a covering of Euclidean space
if
\begin{equation*}
   \RR^n \subseteq \Lambda + \cB(r).
\end{equation*}
The covering radius is the smallest radius $r$ such that
$ \Lambda + \cB(r)$ is a covering, i.e.,
\begin{equation*}
   r_\Lambda^\mathrm{cov} = \min\left\{ r \, : \,
        \Lambda + \cB(r) \text{ is a covering} \right\}.
\end{equation*}
A sequence of lattices $\{ \Lambda_n \}$ is said to be ``good for covering'' if
\begin{equation*}
   \liminf_{n \to \infty}
      \frac{r_{\Lambda_n}^\mathrm{cov}}{r_{\Lambda_n}^\mathrm{eff}} = 1.
\end{equation*}

We now summarize the coding scheme of \cite{UriMLANPaper04}.  In
particular, with $\bu$ denoting a random variable (dither) that is
uniformly distributed over $\cV$, i.e., $\bu\sim \mathrm{Unif}(\cV)$,
we have:
\begin{itemize}
\item
\emph{Transmitter:}
The input alphabet is restricted to $\cV$.  For any $\bv \in \cV$, the
encoder sends
\begin{equation} \label{encoder_output2}
\bx= [\bv -\bu] \bmod \Lambda.
\end{equation}
\item
\emph{Receiver:}
The receiver computes
\begin{equation} \label{decoder_output2}
{\by}'=\left[ \by +\frac{1}{\alpha} \cdot \bu \right] \bmod \Lambda/\alpha
\quad \text{where } \; 0 < \alpha \le 1.
\end{equation}
\end{itemize}

The resulting channel is described by the following lemma
\cite{UriCancel00}.
\begin{lemma}
The channel from $\bv$ to ${\by}'$ defined by
\eqref{AWGNchannel},\eqref{encoder_output2} and
\eqref{decoder_output2} is equivalent in distribution to the
mod-$\Lambda$ channel
\begin{equation} \label{MLAN}
\by'=\left[ \frac{1}{\alpha} \cdot \bv+\bz_\mathrm{eff}'
\right] \bmod \Lambda/\alpha
\end{equation}
with
\begin{equation} \label{eqn:bz_prime}
{\bz}_\mathrm{eff}' =
 \frac{1-\alpha}{\alpha} \cdot \bu + \bz
\end{equation}
\end{lemma}
Note the similarity of the mod-$\Lambda$ channel, depicted in
Fig.~\ref{fig:mlan}, to the equivalent channel representation
\eqref{eqn:modified_scheme} for transmission of a codeword from a
spherical code as depicted in Fig.~\ref{fig:lemma}.

The capacity of the mod-$\Lambda$ channel (\ref{MLAN}) is characterized by
the following theorem.\footnote{This particular form of the theorem is
due to Forney \cite{ForneyMMSE03}.}

\begin{prop}[\cite{UriMLANPaper04,ForneyMMSE03}]
The capacity $C(\Lambda,\alpha)$ of the mod-$\Lambda$ transmission
system is lower bounded by
\begin{equation*}
 C(\Lambda,\alpha) \ge C - \frac{1}{2} \log 2 \pi e
 G(\Lambda) - \frac{1}{2} \log \frac{
 \overline{e}_\alpha}{ \overline{e}_\mathrm{MMSE}}
\end{equation*}
where $\overline{e}_\alpha$ and $\overline{e}_\mathrm{MMSE}$ are the
expected estimation error per dimension using the linear estimator
$\hat{X} = \alpha \cdot X$ and the MMSE estimate $\hat{X}_\mathrm{MMSE} =
\alpha_\mathrm{MMSE} \cdot X$, respectively, where $\alpha_\mathrm{MMSE} =
\SNR/(1+\SNR)$.
\end{prop}
Thus, the gap to capacity may be made arbitrarily small by taking a
lattice $\Lambda$ such that $G(\Lambda)$ is sufficiently close to
${1}/{(2 \pi e)}$ and $\alpha = \alpha_\mathrm{MMSE}$.  That is, to
achieve capacity it is sufficient for $\Lambda$ to be good for
quantization and for $\alpha = \alpha_\mathrm{MMSE}$.  However, as
previously noted, the error exponent is more sensitive to the input
distribution.  In particular, it is no longer sufficient for $\Lambda$
to be good for quantization (as was sufficient to achieve capacity).
We require the additional condition that $\Lambda$ is good for
covering, i.e., that $r^\mathrm{cov}_\Lambda/r^\mathrm{eff}_\Lambda
\rightarrow 1$ as $n \rightarrow \infty$.  Furthermore, as shown in
\cite{LiuMLAN}, a scaling that is strictly less than
$\alpha_\mathrm{MMSE}$ in order to achieve the error exponent.  In
this direction we define the error exponent for the mod-$\Lambda$
channel using a scaling $\alpha$ as
\begin{equation*}
   E_\Lambda(R,\alpha)
 = \limsup_{n \to \infty} \frac{ - \log {\Prm_{e,\Lambda}(n,R,
                                     \alpha)} }{n},
\end{equation*}
where ${\Prm_{e,\Lambda}(n,R,\alpha)}$ is the minimal value of
the average probability of error, ${\Prm_{e,\Lambda}(\cC)}$, over
all $(n,R)$ codes using a scaling $\alpha$, and, in turn, where
$\Prmb_{e,\Lambda}(\cC)$ is the average error
probability of a given $(n,R)$ lattice code averaged over all
codewords.

The error exponent for the mod-$\Lambda$ (just as for the AWGN
channel) is only known for a range of rates.   However, it is shown in
\cite{LiuMLAN} that the error exponent for the mod-$\Lambda$ channel
achieves the random coding exponent, and the scaling $\alpha$ that
achieves this exponent was explicitly found in \cite{LiuMLAN}.
Indeed, as previously noted it is precisely $\asphere^*(R)$
[cf.~\eqref{eqn:alpha_star_sphere}] which we have shown corresponds
to valid spherical regions in Gallager's bound
\eqref{eqn:gallager_region}.   We now provide an intuitive explanation
for this result, which was a question left open by \cite{LiuMLAN}.

We begin by noting that the noise $\bz_\mathrm{eff}'$ in
\eqref{eqn:bz_prime} looks very much like the effective noise
appearing in \eqref{eqn:key}.   Additionally note that the random
vectors $\bb$ and $\bu$ have the same second-moment but while
$\bb$ is spherical, $\bu$ is uniform over $\cV$.
The following proposition makes this notion precise.
\begin{prop}[\cite{UriMLANPaper04}]
\label{prop:voron_ball}
Let $\Lambda$ be any $n$-dimensional lattice that is good for
quantization and covering such that $\sigma^2(\Lambda) = nP$.
Now, consider the random variable $\bu$ that is chosen uniformly from
the fundamental region $\cV(\Lambda)$ and the random variable $\bb$
chosen uniformly from the surface of the ball $\cB(\sqrt{nP})$.   Then
\begin{equation*}
   \log p_{\bu}( \bx ) = \log p_{\bb}( \bx ) + o(1)
\end{equation*}
where $p_{\bu}( \bx )$ and $p_{\bb}( \bx )$ are the probability
density functions of $\bu$ and $\bb$ respectively.
\end{prop}

Define the following modified channel that replaces the
self-noise in \eqref{eqn:bz_prime} with a spherical noise:
\begin{equation} \label{MLANfake}
\by''=\left[ \frac{1}{\alpha} \cdot {\bv}+\bz_\mathrm{eff}''
\right]
\bmod \Lambda/\alpha
\quad \text{with} \quad
\bz_\mathrm{eff}''=
 \frac{1-\alpha}{\alpha} \cdot \bb + \bz.
\end{equation}
Then it follows from Proposition~\ref{prop:voron_ball} that the
error exponent of the original channel \eqref{MLAN} is no worse (in
an exponential sense)
than that of the modified channel \eqref{MLANfake}.
We now bound the error exponent of the channel \eqref{MLANfake} by
using Gallager's technique \eqref{eqn:gallager_region} as before.

It is conceptually much easier if we first consider how one may
remove the $\mod \Lambda/\alpha$ operation appearing in
\eqref{MLANfake}.  In this direction let
\begin{equation*}
 \tilde\by = \frac{1}{\alpha} \cdot {\bv}+\bz_\mathrm{eff}''.
\end{equation*}
Then, following in the footsteps of \cite{LiuMLAN}, we may upper bound the
probability of error by using a suboptimal decoder which first
performs Euclidean distance decoding in the extended codebook
\begin{equation*}
   \cC^\Lambda \myDef \cC + \Lambda
      = \bigcup_{\bc \in \cC} \left\{ \bc + \Lambda
      \right\},
\end{equation*}
and then reduces the result mod-$\Lambda$ to obtain the coset
leader.  We call this sub-optimal decoder the \emph{closest coset
  decoder}.   In other words, the decoder searches for the coset with
minimum Euclidean distance to the received vector and an error occurs
whenever the closest codeword to the received vector does not belong
to the coset of the transmitted codeword (we also take equality as an
error).  That is, using the closest coset decoder an error occurs when
the event
\begin{equation*}
   \cE_\alpha(\tilde{\by},\bc_e) \myDef \left\| \frac{\bc_e}{\alpha} - \tilde{\by}
  \right\|
  \leq
   \left\| \frac{\bc}{\alpha} - \tilde{\by}
  \right\|
\end{equation*}
occurs for some ${\bc_e}$.   Thus, the closest coset decoder satisfies
\begin{equation*}
\begin{aligned}
&\text{decoding rule:}  \quad
   \hat{\bc} = \left[\arg\min_{ \bc  \in \cC^\Lambda }
                    \left\| \frac{\bc}{\alpha} - \tilde{\by} \right\| \right]
\bmod \Lambda/\alpha \\
 & \\
&\text{error event:}  \quad
 \cE_\alpha(\tilde{\by},\bc_e)
\text{ some } \bc_e \in \cC^\Lambda \setminus \{\bc+\Lambda\}
\end{aligned}
\end{equation*}
We may further upper bound the probability of error by using a still
simpler decoder that does not perform coset decoding but rather
searches for the codeword with the minimum Euclidean distance in the
extended codebook.  Thus, with this decoding rule, an error results
even if the closest (to the received vector) codeword in the extended
code belongs to the same coset as the transmitted codeword.  We call
this sub-optimal decoder the \emph{Euclidean distance decoder}.  The
Euclidean distance decoder satisfies
\begin{equation*}
\begin{aligned}
&\text{decoding rule:}   \quad
   \hat{\bc} = \arg\min_{ \bc  \in \cC^\Lambda }
                \left\| \frac{\bc}{\alpha} - \tilde{\by} \right\| \\
 & \\
&\text{error event:}   \quad
  \cE_\alpha(\tilde{\by},\bc_e)
\text{ some } \bc_e \in \cC^\Lambda \setminus \{\bc\}
\end{aligned}
\end{equation*}

Now, by replacing the self-noise by spherical noise and using a
Euclidean distance decoder, we have the
following upper bound on the probability of error in a mod-$\Lambda$
channel:
\begin{align}
\Prm_e^\lambda({\bc})
    &\exple   \pr\left( \,
                 \text{error} \, , \, \frac{\bc}{\alpha} + \bz_\mathrm{eff}''
                   \in \cR(\bc) \right) \notag \\ &\qquad{}  +
                      \pr\left( \,
                \text{error} \, , \,  \frac{\bc}{\alpha} + \bz_\mathrm{eff}''
                    \not\in \cR(\bc) \right) \notag \\
                &\le   \pr\left( \,
                 \text{error} \, , \, \frac{\bc}{\alpha} + \bz_\mathrm{eff}''
                   \in \cR(\bc) \right) \notag \\ &\qquad{}  +
                      \pr\left( \frac{\bc}{\alpha} + \bz_\mathrm{eff}''
                    \not\in \cR(\bc) \right) \notag \\
                 &\le  \sum_{
                         \bc_e \in \cC^\Lambda \setminus
                         \{\bc\}
                           } \pr\left(  \cE_\alpha(\tilde{\by},\bc_e)
                     , \frac{\bc}{\alpha} + \bz_\mathrm{eff}'' \in \cR(\bc)
                      \right) \label{eqn:lattice_prob_bound} \\
                &\qquad {} + \pr\left( \frac{\bc}{\alpha} + \bz_\mathrm{eff}''
                    \not\in \cR(\bc) \right) \notag \\
            &\myDef  \Prm^{\lambda,r}_\mathrm{union}({\bc})
                         + \Prmr^\lambda({\bc})
 \label{eqn:gallager_region_lattice}
\end{align}
where $\Prmr^\lambda({\bc})$ is the probability that
the received vector is not in the region $\cR(\bc)$ and $\Prm^{\lambda,r}_\mathrm{union}({\bc})$ is the sum appearing in
\eqref{eqn:lattice_prob_bound}.  As before, we consider congruent
regions for all $\bc$ and thus simply write $\cR(\bc)$ as
$\cR$ and $\Prmr^\lambda({\bc})$ as $\Prmr^\lambda$.  As done for the AWGN channel, we use
random coding arguments to bound the probability of error.   In this
direction, we denote the ensemble of codes for which the codewords are
drawn uniformly from the Voronoi of a lattice $\Lambda \subset
\RR^n$ as $\hat{\Omega}^{(\Lambda,n)}_0$ and write any
codebook drawn from $\hat{\Omega}^{(\Lambda,n)}_0$ as $\cC$.
Averaging over all the codewords in the code and the ensemble
$\hat{\Omega}^{(\Lambda,n)}_0$ yields
\begin{equation}  \label{eqn:average_region_lattice}
   \Prmb_e^\lambda
   \le \Prmb^{\lambda,r}_\mathrm{union}({\bc})
                         + \Prmr^\lambda({\bc})
   = \Prmb^{\lambda,r}_\mathrm{union}
                         + \Prmr^\lambda.
\end{equation}

As before, begin by considering $\Prmr^\lambda$.   We
have a choice over which region we choose to use in
\eqref{eqn:gallager_region_lattice}.   Note that by taking
$\alpha=\asphere^*(R)$, the effective noise $\bz_\mathrm{eff}''$ is
precisely the same as that found in \eqref{eqn:zeff_def} in
Section~\ref{sec:sphere_region}.   That is the effective noise
$\bz_\mathrm{eff}''$ is the sum of a spherical noise and a Gaussian.
Thus, taking the region $\cR=\Rsphere(\theta(R))$ one
has that
\begin{equation*}
   \Prmr^\lambda = \Prmr
\end{equation*}
where $\Prmr$ was defined via
\eqref{eqn:gallager_region} and the choice of region is left implicit.
Further, as a consequence of Proposition \ref{prop:voron_ball} we can
replace $\bz_\mathrm{eff}''$ with $\bz_\mathrm{eff}'$ and obtain an
asymptotic equality.

This yields the following lemma, establishing
that the sphere-packing error exponent is exponentially equal to the
probability that a random vector that is uniform over the fundamental
region $\cV(\Lambda_n)$ plus a Gaussian vector with independent
identically distributed components of variance $P/\SNR$ remains in a
sphere of radius $\sqrt{n P}/\asphere^*(R) \exp(-R)$ about the scaled
codeword.
\begin{lemma}
\label{lem:key_lem_lattice}
Let $\{\Lambda_n\}$ be a sequence of lattices that is good for
covering and quantization such that $\sigma^2(\Lambda_n) = nP$.
Then
\begin{equation*}
      \exp( -n E_\mathrm{sp}(R;\SNR) ) \expeq
         \pr\left(  \bz_\mathrm{eff}'
             \not\in \cB\left( \frac{\sqrt{n P}}{\asphere^*(R)}
                    \sin \theta(R) \right)
         \right),
\end{equation*}
where
\begin{equation*}
         \bz_\mathrm{eff}' = \frac{1-\asphere^*(R)}{\asphere^*(R)}\bu +
                \bz,
\end{equation*}
and in turn, where $\bu$ is the random variable that is uniform over
$\cV$.
\end{lemma}

Thus, \eqref{eqn:prop_1} holds for the mod-$\Lambda$ channel using the region
and scaling that corresponds to a valid sphere in the AWGN channel.  We
note that using the distance distribution of a random ensemble of
lattice codes it was shown in \cite{LiuMLAN} that \eqref{eqn:prop_2} holds for
the mod-$\Lambda$ channel.   This leads to the following theorem of
\cite{LiuMLAN}.

\begin{prop}[\cite{LiuMLAN}]
If $\Rcrit \le R \le C$, then there exists an $\alpha(R)$ such
that $0 \le \alpha(R) \le \alpha_\mathrm{MMSE}$ and for mod-$\Lambda$
transmission
\begin{equation*}
              \exp\left( -n E_\Lambda(R,\alpha(R)) \right)
                \expeq \exp\left(-n  E_\mathrm{sp}(R;\SNR)\right).
\end{equation*}
Furthermore, the exponent $E_\mathrm{sp}(R;\SNR)$ can be achieved with
Euclidean decoding.
\label{thrm:Liu}
\end{prop}

\begin{figure*}
\begin{center}
\scalebox{1}{
   \begin{psfrags}
      \psfrag{A}[bc]{dither}
      \psfrag{B}[bc]{($\bu$ or $\bb$)}
      \psfrag{C}[br]{$\frac{\bc}{\alpha}$}
      \psfrag{D}[tr]{$\bx$}
      \psfrag{E}[bl]{$\cV$}
      \psfrag{F}[cl]{spherical code}
      \psfrag{G}[cr]{spherical code}
      \psfrag{H}[cr]{inflated}
      \psfrag{R}[cc]{\Huge $\star$}
      \psfrag{O}[cc]{\Large $\circledcirc$}
      \includegraphics[scale=0.55]{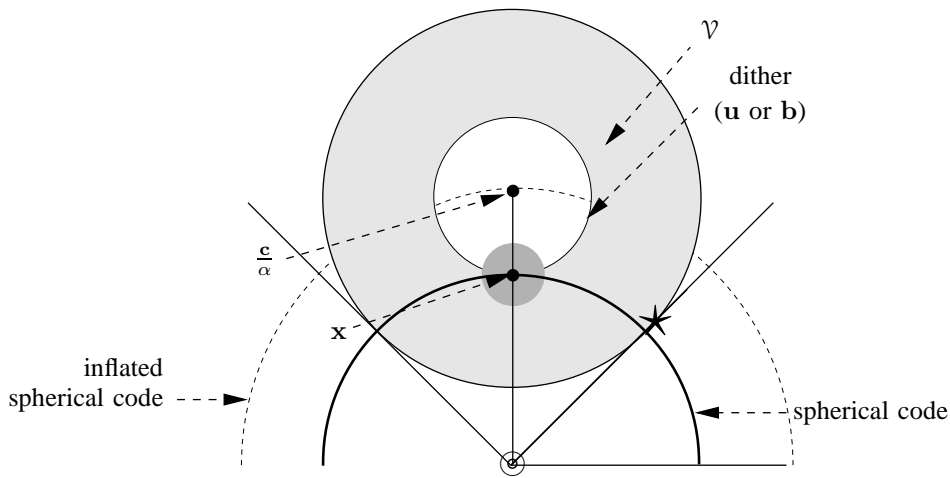}
    \end{psfrags}
}
\end{center}
  \caption{A geometric relationship between the typical error events
    for a spherical code and the mod-$\Lambda$ transmission scheme.  A
    random dither $\bb$ (resp.  $\bu$) is added to the scaled code word
    $\bc$ (resp.  $\bc$) and $\bx$ is transmitted.  The common typical
    error event is depicted by {\Large $\star$}.}
    \label{fig:relation_fig}
\end{figure*}

Reexamining Fig.~\ref{fig:sphere_gauss} of Section
\ref{sec:sphere_region} we can see why both spherical codes and the
mod-$\Lambda$ transmission scheme may achieve the sphere-packing error
exponent.   By identifying the transmitted codewords $\bx$, as depicted
in Fig.~\ref{fig:relation_fig}, we can interpret the center of the
spherical region, $\bc/\alpha$, as the selected codeword in the
mod-$\Lambda$ channel and $\Rsphere(\theta(R))$ as a spherical
approximation to the Voronoi region of the codeword.

We can extend our analysis to rates less than the critical rate.   For
completeness we present a separate proof of this in
Section~\ref{sec:second_prelim} which shows that not only are the
error exponents in the AWGN and mod-$\Lambda$ channels equal, but the
typical error events coincide.   In the following section we summarize
our characterization of the AWGN error exponent for rates above and
below the critical rate.

\section{The AWGN Error Exponents for Low Rates}
   \label{sec:second_prelim}

For rates less than the critical rate the best known lower bound is
the maximum of the \emph{random coding error exponent},
$E^r_\mathrm{AWGN}(R;\SNR)$, and the \emph{expurgated error exponent},
$E^x_\mathrm{AWGN}(R;\SNR)$.  As Fig.~\ref{fig:awgn_exp} reflects, the
random coding error exponent is the larger error exponent for all
rates greater than
\begin{equation*}
   R_x = \frac{1}{2} \log \left( \frac{1}{2} \left( 1 + \sqrt{1+\frac{\SNR}{4}}
                           \right) \right)
\end{equation*}
and is equal to the sphere-packing error exponent for rates greater
than the critical rate.  Hence, the sphere-packing error exponent is
tight for all rates greater than the critical rate.  The error
exponent of the AWGN channel, however, is still not known for all
rates.  Recent progress has been made to show that the random coding
error exponent is indeed the correct error exponent for a range of
rates less than the critical rate; see \cite{BurnRel07} and references
therein.  For all rates less than the critical rate the random coding
error exponent is linear.  More precisely, the random coding error
exponent is
\begin{equation}
   E^r_\mathrm{AWGN}(R;\SNR) \myDef \left\{
      \begin{array}{lc}
       E_\mathrm{r}(R;\SNR)        & \text{ if } \; 0 \le R \le \Rcrit \\
       E_\mathrm{sp}(R;\SNR) & \text{ if } \; \Rcrit < R \le C
      \end{array}
                         \right.
\label{eqn:random_exp}
\end{equation}
where
\begin{equation*}
   E_\mathrm{r}(R;\SNR) = E_\mathrm{G}(\beta_G',1; \SNR)
\end{equation*}
and, in turn,
\begin{equation} \label{eqn:g_beta}
  \beta_G' = \frac{1}{2} \left(1 + \frac{\SNR}{2} +
                 \sqrt{ 1 + \frac{\SNR^2}{4}} \right).
\end{equation}
Recall that $E_\mathrm{G}(\beta,\rho; \SNR)$ was defined in \eqref{eqn:gall_Eg}
and note that $\beta_G'$ is independent of the rate.

For all rates less than $R_x$ the expurgated error exponent is greater
than the random coding error exponent.   In order to precisely define
the expurgated error exponent, recall that the \emph{minimum distance}
of a code is the smallest distance between any two codewords in a
code.   The expurgated error exponent geometrically corresponds to
errors occurring between the closest two codewords of a code that
achieves the best known minimum distance (as this is the dominating
error event at low rates; see, e.g., \cite{BargForney}).  Conversely,
by using an upper bound on the minimum distance one can arrive at the
\emph{minimum distance} upper bound on the error exponent.   These error
exponents are apparent in Fig.~\ref{fig:awgn_exp}.  The expurgated
error exponent\footnote{ This error exponent may be achieved by
  drawing a uniform code over the sphere and then expurgating all
  codewords that fall within a distance equal to the best known
  minimum distance of any other codeword.} is
\begin{equation*}
   E^x_\mathrm{AWGN}(R;\SNR) \myDef
\frac{\SNR}{4} \left(1 - \sqrt{ 1 - \exp( -2 R ) }
               \right).
\end{equation*}
Thus,
the best known lower bound on the error exponent of the AWGN channel is
\begin{equation*}
   E_\mathrm{AWGN}(R;\SNR)
   \myDef \left\{
      \begin{array}{ll}
       E^x_\mathrm{AWGN}(R;\SNR) & \text{ if } \; 0 \le R \le R_{x} \\
       E^r_\mathrm{AWGN}(R;\SNR) & \text{ if } \; R_{x} < R \le C
      \end{array}
                         \right.
\end{equation*}

In the preceding sections we provided a simple proof that, for a
variety of regions $\cR$, one may achieve the sphere-packing
bound using Gallager's bounding technique for $R > \Rcrit$ under
the assumption that \eqref{eqn:prop_2} holds and used this to provide a simple
explanation for the error exponent of the mod-$\Lambda$ channel.   In
the following section we prove that \eqref{eqn:prop_2} does indeed hold,
provide exponential bounds for $\Prmub^r$ and
show that this bound is exponentially equal to $E_\mathrm{AWGN}(R;\SNR)$.

\section{Geometric Derivation of the Random Coding and
Expurgated Error Exponents}
\label{sec:low_rates}

Begin by recalling from \eqref{eqn:gall_bound} that
$\Prmub^r$ is a union bound over pairwise
errors averaged
over the random ensembles of spherical codes $\Omega^{(n)}_0$.   While
this ensemble is sufficient to achieve the random coding error
exponent we require a more general ensemble of codes to derive the
best known bound on the error exponent for the AWGN and mod-$\Lambda$
channels as well as provide the final geometric link between the error
exponents of these two channels.  In this direction let
$\Omega^{(n)}_d(R) =(\Omega^{(n)}_0,d_\Omega(R))$ be the ensemble of
rate $R$ random spherical codes where expurgation has been applied
such that the minimum distance is $d_\Omega(R)$.  Note, with this
notation $\Omega^{(n)}_0(R)$ is the ensemble with minimum distance $0$
or the random spherical ensemble $\Omega^{(n)}_0$ that was introduced
previously in Section \ref{sec:geom_awgn}.   Moreover, it is known from
\cite{ShannErrorProb} that no rate loss in incurred from the
expurgation process for any $d_\Omega(R)$ such that
\begin{equation*}
  d_\Omega(R) \le d_\mathrm{min}(R),
\end{equation*}
where
\begin{equation} \label{eqn:dmin_eq}
   d_\mathrm{min}(R) \myDef \sqrt{2 - 2 \sqrt{ 1 - \exp( -2 R ) } }.
\end{equation}
Henceforth we consider ensembles of codes such that
$d_\Omega(R) \le d_\mathrm{min}(R).$ In particular, we consider the
ensembles
\begin{align}
      \Omega^{(n)}_\mathrm{I}(R)   &= (\Omega^{(n)}_0,0)
        \label{eqn:e_I} \\
      \Omega^{(n)}_\mathrm{II}(R)  &= (\Omega^{(n)}_0, e^{-R})
        \label{eqn:e_II} \\
      \Omega^{(n)}_\mathrm{III}(R) &= (\Omega^{(n)}_0, d_\mathrm{min}(R))
        \label{eqn:e_III}
\end{align}
and denote exponent of the average probability of error for these
ensembles as $E_\mathrm{I}(R;\SNR)$, $E_\mathrm{II}(R;\SNR)$ and
$E_\mathrm{III}(R;\SNR)$ respectively.   We show in the sequel that
\begin{equation*}
\exp( -n E_\mathrm{I}(R;\SNR) ) \exple \exp( -n E^r_\mathrm{AWGN}(R;\SNR))
\end{equation*}
and
\begin{equation*}
\exp( -n E_\mathrm{III}(R;\SNR) ) \exple \exp( -n E_\mathrm{AWGN}(R;\SNR)).
\end{equation*}
Moreover, in Section \ref{sec:lattice} we show that the mod-$\Lambda$
channel can obtain an average probability of error that is
exponentially equal to that obtained by the ensemble
$\Omega^{(n)}_\mathrm{II}(R)$.   We begin by examining the exponential
behavior of the probability of having a pairwise error while remaining
inside the cone $\Rcone(\theta)$ for the ensemble
$\Omega^{(n)}_\mathrm{I}(R)$.  That is, the exponent of
$\Prmb^r_\mathrm{union}$ for the code ensemble with no minimum distance
constraint.

\begin{figure*}
\begin{center}
   \begin{tabular}{ccc}
      \begin{psfrags}
      \psfrag{M}[rb]{$\sqrt{nP} \be_y \,$}
      \psfrag{ME}[lt]{$\bc_e$}
      \psfrag{X}[tc]{$x_c$}
      \psfrag{Y}[cc]{$y_c$}
      \psfrag{B}[cc]{$\beta \sqrt{nP} $}
      \psfrag{TA}[rc]{$\thetaD{d}$}
      \psfrag{TD}[rc]{$\theta(R)$}
      \psfrag{D}[lc]{$\cD_c(d)$}
      \psfrag{D2}[cc]{$\frac{d \sqrt{nP}}{2}$}
      \includegraphics[scale=0.592]{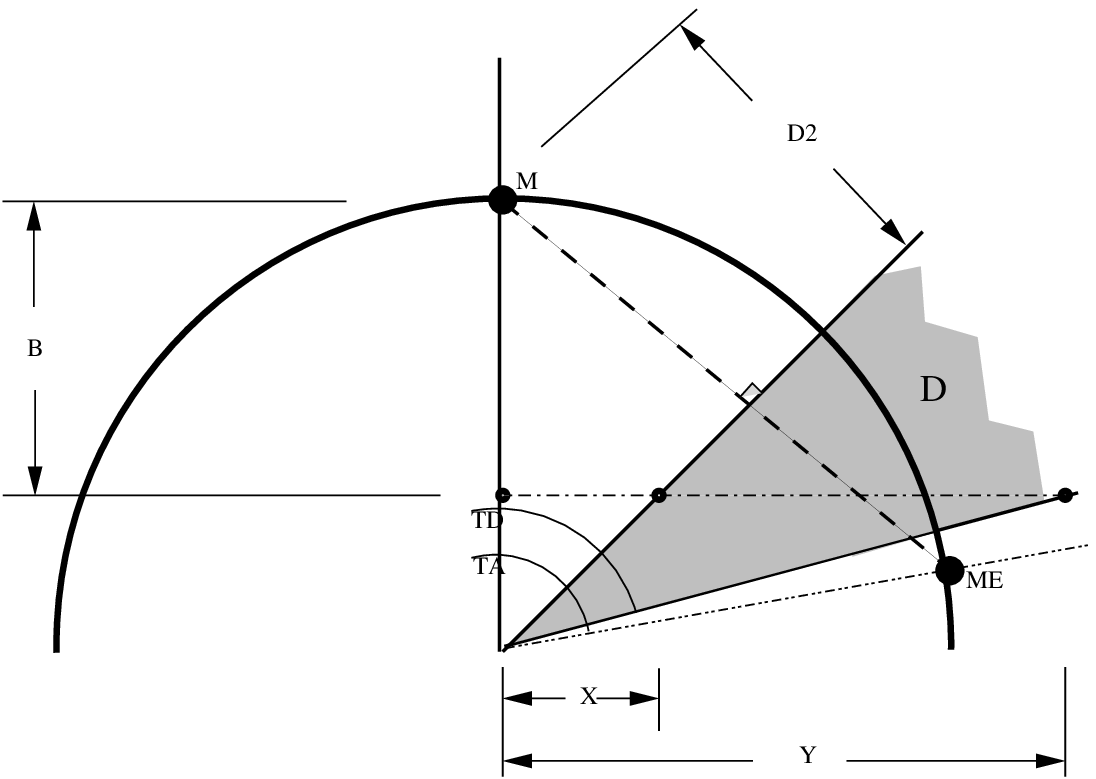}
      \end{psfrags}
   & \quad  &
      \begin{psfrags}
      \psfrag{TA}[rc]{\small $\theta(R)$}
      \psfrag{TD}[rc]{\small $\thetaD{d}$}
      \psfrag{D}[lc]{$\cD(d,\theta,\beta)$}
      \includegraphics[scale=1.72]{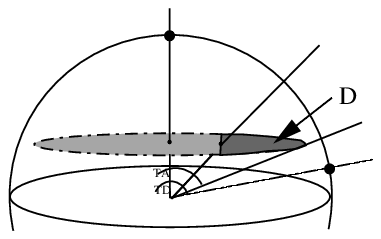}
      \end{psfrags}
   \\
      (a) & & (b)
   \end{tabular}
   \caption{The parameters for the derivation of the AWGN error
     exponent.   (a) A 2D representation of the bounding technique.
     The region $\cD_c(d)$ corresponding to an error with a
     codeword at distance $d$ condition on the event that the noise
     remains in $\cR$ can be seen shaded in gray.  (b) A three
     dimensional representation of the region corresponding to an
     error with a codeword at distance $d$ condition on the event that
     the noise remains in $\cR$ and the radial component of the
     noise.}
   \label{fig:big_picture_awgn}
\end{center}
\end{figure*}

Recall that conditioned on the event that the sum of the codeword and
the noise remains inside $\Rcone(\theta)$ an error occurs
between a codeword, say $\bc_e$, at a distance $d$ if the codeword
plus the noise crosses the ML plane between $\bc$ and $\bc_e$.   We let
$\cD_c(d,\theta)$ be the region corresponding to this event.
That is, we let $\cD_c(d,\theta)$ be the intersection of the
cone $\Rcone(\theta)$ with the half space that orthogonally
bisects a cord of length $d$ that has one end point at the transmitted
codeword and the other end at $\bc_e$.   This can be seen as the shaded
region in Fig.~\ref{fig:big_picture_awgn}.   We write
$\cD_c(d)$ for simplicity.

Similar to the previous sections we find the typical error events or
the distance, $d$, and $\beta$ that maximize $\Prmb^r_\mathrm{union}$.
More precisely for each $d$ we find the typical $\beta$ and then find
the typical $d$.  It is often simpler to consider the angle made
between the transmitted codeword and any codeword at a distance $d$
instead of the distance itself and denote this angle by $\thetaD{d}$,
i.e.,
\begin{equation} \label{eqn:thetad_def}
   \thetaD{d} \myDef  2 \arcsin\left(d/2\right)
              = \arccos\left( 1 - {d^2}/{2} \right).
\end{equation}

In Appendix \ref{append:dist_dist} it is shown that
\begin{align}
    &\Prmub^r
    \le 2 K  \max_{ 0 \le d \le 2 }
    \pr\left( \sqrt{nP} \cdot \be_y + \bz \in \cD_c(d)
     \right)  \notag \\
    & \qquad \qquad{} \times \exp{\left(n R + (n-1) \log
    \left( d \sqrt{1 - \frac{d^2}{4} } \right) \right) }
   \label{eqn:cone_bound}
\end{align}
where $K$ is a normalizing constant.  In order to form an exponential
bound for \eqref{eqn:cone_bound} we begin by examining the exponential
behavior of $\pr\left( \sqrt{nP} \cdot\be_y + \bz \in \cD_c(d)
\right)$, i.e., the probability of having a pairwise error with a
codeword at a distance $d$ while remaining inside the cone
$\Rcone(\theta)$.  We again use the tangential sphere bound and thus we
let $\cD_c(d,\theta,\beta)$ be the intersection of $\cD_c(d,\theta)$
with the hyperplane, say $\cH$, at a distance of $\beta \sqrt{n P}$
from the transmitted codeword.  More precisely, let $\cH$ be the
hyperplane such that $\be_y' \bx = \beta \sqrt{nP}$ for all $\bx \in
\cH$.  The $n-1$ dimensional region $\cD_c(d,\theta,\beta) =
\cD_c(d,\theta) \cap \cH$ and can be seen in
Fig.~\ref{fig:big_picture_awgn}.

Integrating along the radial component of the noise we have
\begin{align}
 &
 \pr\left(  \sqrt{nP} \cdot \be_y + \bz \in \cD_c(d) \right)
   \notag \\
 &= \int_{-\infty}^{\infty} e^{ -n \frac{\SNR \bcone^2 }{2} }
          \pr\left(\sqrt{nP} \cdot \be_y  + \bz
             \in \cD_c(d,\theta,\beta) \right) d\bcone  \notag\\
 &= \int_{-\infty}^{\infty} e^{ -n \frac{\SNR \bcone^2 }{2} }
          \pr\left(z_2 \ge x_c ,
                   \sum_{i=2}^n z_i^2 \le y_c^2 \right) d\bcone
\label{eqn:int_bound_line}
\end{align}
where
\begin{equation*}
   x_c = x_c(\bcone,\thetaD{d}) =  \sqrt{n P} (1+\bcone) \tan
   \frac{\thetaD{d}}{2}
\end{equation*}
and
\begin{equation*}
y_c = y_c(\bcone) = \sqrt{n P} (1+\bcone) \tan \theta(R)
\end{equation*}
can be derived through the geometry in
Fig.~\ref{fig:big_picture_awgn}.  That is, when the radial component
of the noise has magnitude $\bcone$, the probability $\pr\left(
\sqrt{nP} \cdot \be_y + \bz \in \cD_c(d) \right)$ is, geometrically
speaking, simply the probability that the second component of the
noise is greater than $x_c$ (so that the codeword plus the noise is in
the decoding region for a different codeword) while the magnitude of
the second thorough $n$th component of the noise is less than $y_c$
(so that the codeword plus the noise is in $\Rcone(\theta)$).  Before
proceeding, we recall the following bound on Gaussian vectors.

\begin{prop}[\cite{PoltyrevUncon}]
\label{prop:chernoff_constrain}
Let $z_1, z_2, \ldots, z_n$ be {\it i.i.d} zero-mean Gaussian random
variables with variance $\sigma^2$.  Let $\tilde{\bz} = ( z_1, \ldots,
z_n)$.   Then, if $n \ge 2$,
\begin{equation*}
\pr( |z_1|  \ge \sqrt{n P} x ,  \| \tilde{\bz} \|  \le \sqrt{n P} y )
 \le e^{-n \, \Ebdt\left(x,y ; \frac{P}{\sigma^2} \right)}
\end{equation*}
where the exponent $\Ebdt$ is defined via
\begin{equation}
   2 \, {\Ebdt}\left(x,y ; \tau \right)
   \myDef  \left\{
        \begin{array}{cc}
          {\tau  x^2  } &
             \text{if $y^2 - x^2 \ge \frac{1}{\tau}$ } \\
           {\tau} y^2  -
             \log( e  \tau ( y^2 - x^2))
                 & \text{otherwise}
        \end{array}
    \right.
    \label{eqn:expon_Ebd_tilde}
\end{equation}
\end{prop}

Note that in \eqref{eqn:int_bound_line} we may bound the integral of
the right-hand side by one times the largest term.  Applying
Proposition \ref{prop:chernoff_constrain} we have that the probability
that the received vector is outside the cone $\Rcone(\theta)$
satisfies
\begin{align}
&\pr\left( \sqrt{n P} \cdot \be_y + \bz \in \cD_c(d) \right)
 \notag \\
 &\quad \exple \max_{  \bcone  }
          \exp\bigl( -n {\Ebd}\left(\bcone,x_c(\bcone,\thetaD{d}),y_c(\bcone);
                     \SNR \right) \bigr)
 \label{eqn:cone_exponent}
\end{align}
where
\begin{equation}
   2 \, {{\Ebd}}\left(\beta,x,y ; \tau \right)
   \myDef   \tau \beta^2 +  {\Ebdt}\left(x,y ; \tau \right)
    \label{eqn:expon_Ebd}
\end{equation}
Thus, \eqref{eqn:cone_bound} becomes
\begin{equation}
   \Prmub^r
    \exple
   \max_{  0 \le d \le 2 }
   \max_{  \beta > -1 }  \,
   \exp\bigl( -n \,  E_\mathrm{bnd}(\theta,d,\beta,R;\SNR) \bigr)
   \label{eqn:awgn_bound_final}
\end{equation}
where
\begin{align}
   & E_\mathrm{bnd}(\theta,d,\beta,R;\SNR) \notag \\
   &\quad = \Ebd(\beta,x_c,y_c;\SNR) - \frac{1}{2}  \log{\left[
            d^2 \left(1 - \frac{d^2}{4}\right)
                        \right]} - R.  \label{eqn:Ecrc_def}
\end{align}

It is a simple, yet lengthy, process to find the value of $\beta$ that
maximizes $\Prmub^r$.   We provide a full
derivation of the optimal $\beta$ in Appendix \ref{append:line}, but
for now note that is equal to
\begin{equation}
\bcone^*(\theta,\thetaD{d};\SNR)
  = \left\{
                \begin{array}{lc}
\beta^*(\theta;\SNR)
&
\text{if }  R(\theta) > \Rcrit(\thetaD{d})
\\
                     \cos^2\left( \frac{\thetaD{d}}{2} \right)
                      & \text{ otherwise }
                \end{array}
                \right.
\label{eqn:beta_star_d}
\end{equation}
where $\beta^*(\theta;\SNR)$ was defined in \eqref{eqn:opt_beta_sphere} and
\begin{equation*}
  \Rcrit(\thetaD{d}) \myDef - \log
                       \sqrt{1 - \frac{2 \, \SNR \cos^4 \frac{\thetaD{d}}{2}}{
                         2 + \SNR \, (1 + \cos \thetaD{d}) } }.
\end{equation*}

Note that for a given $\thetaD{d}$ if $R(\theta) >
R_\mathrm{crit}(\thetaD{d})$ then $ \bcone^*$ is independent of $d$
and exactly equal to the optimal $\beta$ in the derivation of the
upper bound for $\Prmr$
[cf.~\eqref{eqn:opt_beta_sphere}].  That is, the typical error event
of $\Prmub^r$ corresponds to the typical error
event for $\Prmr$.  In fact, if this were not the case
for $\theta = \theta(R)$ then we would either be able to improve the
sphere-packing error exponent by making the region $\Rcone(\theta(R))$
larger (if $\Prmr < \Prmub^r$) or
be unable to show that the sphere-packing error exponent is tight (if
$\Prmr < \Prmub^r$).  Thus, the
minimizing distance should be the point that is tangent to
$\Rcone(\theta)$ at $\beta^*(\theta(R);\SNR)$.  By the law of cosines
this would imply that $d_c^*(\theta) = { \sqrt{2} \sin \theta }$ for
$R(\theta) > \Rcrit(\thetaD{d})$.  We show in Appendix
\ref{append:line} that this is indeed the case and the value of $d$
that maximizes $\Prmub^r$ is
\begin{equation} \label{eqn:dc_star}
   d_c^*(\theta) = \left\{
                \begin{array}{lc}
                   \sqrt{2} \sin \theta &
                   \text{if } R(\theta) > \Rcrit \\
                     d_\mathrm{crit}                & \text{ otherwise }
                \end{array}
                \right.
\end{equation}
where\footnote{Note this is equal to $\sqrt{2/\beta_G'}$.}
\begin{equation*}
   d_\mathrm{crit} = \sqrt{ 2 + \frac{4}{\SNR} - 2 \sqrt{1 + \frac{4}{\SNR^2}}}
\end{equation*}
Combining \eqref{eqn:beta_star_d} and \eqref{eqn:dc_star} in to one
equation yields the following definition and lemma.  Let
\begin{equation} \label{eqn:fbdn}
   f_\mathrm{bnd}(d,\theta, R;\SNR) =
   \exp\left( -n  \Et_\mathrm{bnd}\left( d, \theta, R ;
   \SNR \right)
       \right)
\end{equation}
where $\Et_\mathrm{bnd}(d,\theta;\SNR)$ is defined at the
bottom of the page in \eqref{eqn:tilde_bnd}.
\begin{figure*}[b]
\hrulefill
\begin{equation}
   \Et_\mathrm{bnd}(d,\theta;\SNR) = \left\{
                \begin{array}{lc}
                    {E_\mathrm{sp}}\left( R(\theta) ; \SNR \right)
                    - \log \left( \sin \theta  \right) - R
                   & \text{if } R(\theta) > \Rcrit(\thetaD{d}) \\
                    {\SNR}/{8} \cdot d^2
                    - \log\left( d \sqrt{1-d^2/4}  \right) - R
                   & \text{ otherwise }
                \end{array}
                \right.
\label{eqn:tilde_bnd}
\end{equation}
\end{figure*}

This yields the following lemma.

\begin{lemma}
Consider the sequence of ensembles of random spherical codes
$\{\Omega^{(n)}_\mathrm{I}(R)\}$.  Then, if $\cR = \Rcone(\theta(R))$
and $\Rcrit < R < C$,
\begin{equation*}
   \Prmub^r \le
     f_\mathrm{bnd}(d_c^*(\theta(R)),\theta(R);\SNR)
\end{equation*}
\end{lemma}

From the above discussion, it is clear that in the case that $\theta =
\theta(R)$ we have that the typical error events for
$\Prmb^r_\mathrm{union}$ and $\Prmr$ are equivalent
  and
$\Prmub^r \expeq \Prmr$ for
all $R \ge \Rcrit$.  Thus \eqref{eqn:prop_2} holds and we have shown that
the sphere-packing error exponent is indeed a valid lower bound on the
error exponent of the AWGN channel.   We state this in the following
lemma.

\begin{lemma}
Consider the sequence of ensembles of random spherical codes
$\{\Omega^{(n)}_\mathrm{I}(R)\}$.  Then, if $\cR = \Rcone(\theta(R))$, the average probability of error for $\Rcrit
< R < C$ is upper bounded by
\begin{equation*}
  \Prmb_e \exple \exp( -n E_\mathrm{sp}(R;\SNR) ).
\end{equation*}
\end{lemma}

Note that if $R(\theta) < \Rcrit$ both $d_c^*(\theta)$ and
$\bcone^*$ are independent of $\theta$.   Thus, if $\theta =
\theta(R)$ then for all rates $R < \Rcrit$ it is clear that one
may fix $\theta = \pi/2$ and obtain the same result for
$\Prmub^r$ as if one had used $\theta =
\theta(R)$.   In this direction, let
\begin{equation*}
   \theta^r_\mathrm{AWGN}(R) \myDef \left\{
      \begin{array}{lc}
        \pi/2 & \text{ if } 0 \le R \le \Rcrit \\
        \theta(R) & \text{ if } \Rcrit < R \le C \\
      \end{array}
                                 \right.
\end{equation*}
Note that this implies that $\Prmub^r$ must be equal to the union
bound over all codewords, $\Prmub$.  Now we have the following
theorem.

\begin{theorem}
Consider the sequence of ensembles of random spherical codes
$\{\Omega^{(n)}_\mathrm{I}(R)\}$.
If $\cR = \Rcone(\theta^r_\mathrm{AWGN}(R))$, then
Gallager's bounding technique \eqref{eqn:gallager_region}
for the average probability of error
satisfies the following two properties for $0 \le R \le C$:
\begin{equation*}
   \begin{aligned}
      1'.  \qquad &   \Prmub^r \exple
                     \exp\left(-n \,  E^r_\mathrm{AWGN}(R;\SNR)\right)      \\
      2'.  \qquad &  \Prmub^r \expge
                    \Prmr        \\
   \end{aligned}
\end{equation*}
where $ E^r_\mathrm{AWGN}(R;\SNR) $ was defined in \eqref{eqn:random_exp}.
\label{thrm:sl_thrm}
\end{theorem}
A proof is provided in Appendix \ref{append:line}.

Note that the typical error events of Theorem \ref{thrm:sl_thrm}
happen with codewords at a distance $d_\mathrm{crit}$ for rates less than
$\Rcrit$.   Hence, by using a code ensemble such that
$d_\Omega(R) > d_\mathrm{crit}$ one expects to be able to improve upon
our current bound for rates such that $d_{\Omega}(R) > d_\mathrm{crit}$.
For example,  we consider the ensemble of codes $\Omega^{(n)}_\mathrm{III}(R)$.
It is straightforward to check that for the ensemble
$\Omega^{(n)}_\mathrm{III}(R)$
one has $d_\mathrm{min}(R) > d_\mathrm{crit}$ for $R < R_\mathrm{III}$,
where
\begin{equation*}
   R_\mathrm{III} =
   R_x = \frac{1}{2} \log \left( \frac{1}{2} \left( 1 + \sqrt{1+\frac{\SNR}{4}}
                           \right) \right).
\end{equation*}
Additionally, it is easy to see that for
$\theta=\theta^r_\mathrm{AWGN}(R)$ the ensemble
$\Omega^{(n)}_\mathrm{III}(R)$ has typical error events that occur at
a distance $d_c^*(\theta) = d_\mathrm{min}(R)$ if $R <
R_\mathrm{III}$.  Hence, the typical error events occur with codewords
at a distance
\begin{equation}
\label{eqn:d_typ}
d_\mathrm{typ}(R) \myDef \left\{
        \begin{aligned}
              d_\mathrm{min}(R) & \qquad \text{ if } \, 0 \le R \le R_x \\
              d_\mathrm{crit}  &  \qquad \text{ if } \, R_x < R \le \Rcrit \\
           \sqrt{2} \exp(-R) &  \qquad \text{ if } \, \Rcrit < R  \le C \\
        \end{aligned}
        \right.
\end{equation}
This yields the following characterization of the error exponent of the
AWGN channel.
\begin{theorem}
\label{thrm:awgn_char_notheta}
Consider the sequence of ensembles of random spherical codes
$\{\Omega^{(n)}_\mathrm{III}(R)\}$.
If $\cR = \Rcone(\theta^r_\mathrm{AWGN}(R))$, then
the typical error events occur with codewords at a distance
$d_\mathrm{typ}(R)$ and for $0 \le R \le C$:
\begin{align}
      &A.  \quad \Prmub^r \exple
 f_\mathrm{bnd}(d_\mathrm{typ}(R),\theta^r_\mathrm{AWGN}(R);\SNR) \\
      &B.  \quad \Prmr
                   \exple
 f_\mathrm{bnd}(d_\mathrm{typ}(R),\theta^r_\mathrm{AWGN}(R);\SNR)
\end{align}
Moreover,
\begin{equation*}
 f_\mathrm{bnd}(d_\mathrm{typ}(R),\theta^r_\mathrm{AWGN}(R);\SNR)
\expeq
e^{-n \,  E_\mathrm{AWGN}(R;\SNR) }.
\end{equation*}
\end{theorem}

Note, in Theorem \ref{thrm:awgn_char_notheta}, $\Prmr$
is less than our bound on $\Prmub^r$,
$f_\mathrm{bnd}$.  In order to improve the bound on the error exponent
it is natural to ask whether the choice of $\theta =
\theta^r_\mathrm{AWGN}$ in Theorem \ref{thrm:awgn_char_notheta} is the
best choice for all rates $R < \Rcrit$ since there is some slack in
the choice $\theta = \theta^r_\mathrm{AWGN}$ due to the fact that
$\Prmub^r \expge \Prmr$.  We
have shown that for the ensemble $\Omega^{(n)}_\mathrm{I}(R)$ one can
not do better in terms of the average probability of error for the
ensemble using Gallager's technique.  We next show that so long as
$\theta$ is within reason the choice of $\theta$ has no effect on the
resulting bound on the average probability of error.

In order to precisely characterize the freedom one has in choosing
$\theta$, we require the following definitions.   Let, for
$K \ge 1/\SNR$,
\begin{equation*}
   \theta_\zeta(K;\SNR) = \arcsin\left( \sqrt{ 1 - \frac{1}{ K (1+K) \SNR }
                                }\right).
\end{equation*}
Note that with this
parametrization of $\theta_\zeta$ one has a simple parametrization of the
sphere-packing exponent (or alternatively a simple parametrization of the
upper bound $\Prm_{\rm region}$ for the
region $\cR = \Rcone( \theta_\zeta(K;\SNR) )$) in terms of $K$.
More precisely, one has
\begin{align}
  &E_\mathrm{sp}\left( R(  \theta_\zeta(K;\SNR) ) ; \SNR \right) \notag \\
  &\quad = \frac{ -1 + K \SNR
        - K \log\left( 1 + \frac{1}{K} - \frac{1}{K^2 \SNR} \right) }{2 K}.
 \label{eqn:sphere_param}
\end{align}
and
\begin{equation*}
   \Prmr \exple \exp\left( -n \, \frac{ -1 + K \SNR
        - K \log\left( 1 + \frac{1}{K} - \frac{1}{K^2 \SNR} \right) }{2 K}
   \right).
\end{equation*}
Using this parametrization we now precisely characterize the freedom one has
in choosing $\theta$.  We
let, for $R(\theta_\zeta(K;\SNR)) < \Rcrit$,
$z(K;d,R,\SNR)$ be $2 K$ times the difference of $E_\mathrm{sp}$ and
$E_\mathrm{bnd}$
for the region $\cR = \Rcone( \theta_\zeta(K;\SNR) )$ and an ensemble with
typical error events that occur with codewords at a distance
$d_c^* = d$.   More precisely we let
\begin{align}
z(K;d,R,\SNR) &= -1+K \left(2 R+\left(1-\frac{d^2}{4}\right) \SNR\right)
  \notag \\
&\quad{} + K \log\left[\frac{d^2 \left(1-\frac{d^2}{4}\right) K^2 \SNR}{
                       K (1+K) \SNR -1}\right].  \notag
\end{align}

We can now use the characterization of the typical error events
\eqref{eqn:d_typ} to help find the minimal $\theta$ such that one may
obtain the proper error exponent.   First, we state the following
property of the function $z(K;d,R,\SNR)$.
\begin{lemma}
The function $z(K;d,R,\SNR)$, as a function of $K$, has a unique zero
on the interval $[1/\SNR,\infty)$ for $d \ge 0$, $R \ge 0$ and $\SNR >
  0$.
\label{lem:z}
\end{lemma}
A proof is provided in Appendix \ref{sec:z_proof}.

We let $K_\zeta(d;R,\SNR)$ be the unique root of $z(K;d,R,\SNR)$ on the
interval $[1/\SNR,\infty)$.
Hence, for rates $R \le \Rcrit$, if one chooses
$\theta \ge \theta_\zeta(K_\zeta(d;R,\SNR);\SNR)$, then
\begin{equation} \label{eqn:balance}
   \Prmr \exple
   f_\mathrm{bnd}(d_\mathrm{typ}(\theta(R)),\theta(R);\SNR)
\end{equation}
Thus, we let $\theta_\mathrm{AWGN}(R;\SNR)$ be the smallest $\theta$ such that
\eqref{eqn:balance} holds
for the ensemble $\Omega^{(n)}_\mathrm{III}(R)$ using a region
$\cR = \Rcone(\theta)$.  That is, $\theta_\mathrm{AWGN}(R;\SNR)$ satisfies
\eqref{def:theta_awgn} at the bottom of the page.
\begin{figure*}[b]
\hrulefill
\begin{equation} \label{def:theta_awgn}
   \theta_\mathrm{AWGN}(R;\SNR) =
   \left\{
   \begin{aligned}
      \theta_\zeta(K_\zeta(d_\mathrm{typ}(R);R,\SNR);\SNR) &
      \qquad \text{ if } 0 \le R < \Rcrit \\
      \arcsin \exp\left( - R \right) &
      \qquad \text{ if } \Rcrit \le R \le C
   \end{aligned}
   \right.  .
\end{equation}
\end{figure*}
Then, if we use any region
$\cR = \Rcone(\phi(R))$ such that
\begin{equation*}
   \theta_\mathrm{AWGN}(R;\SNR) \le \phi(R) \le \arcsin \exp(-R),
\end{equation*}
the results of Theorem \ref{thrm:awgn_char_notheta} still hold.  We state this
in the following theorem.

\begin{theorem}
\label{thrm:awgn_char_theta}
Consider the sequence of ensembles of random spherical codes
$\{\Omega^{(n)}_\mathrm{III}(R)\}$ and let
$\phi(R)$ be given such that
  $$\theta_\mathrm{AWGN}(R;\SNR) \le \phi(R) \le \arcsin \exp(-R).$$
If $\cR = \Rcone(\phi(R))$, then
the typical error events occur with codewords at a distance
$d_\mathrm{typ}(R)$ and for $0 \le R \le C$:
\begin{align*}
      &A.  \quad \Prmub^r \exple
                     f_\mathrm{bnd}(d_\mathrm{typ}(R),\theta^r_\mathrm{AWGN}(R);\SNR) \\
      &B.  \quad \Prmr
                   \exple f_\mathrm{bnd}(d_\mathrm{typ}(R),\theta^r_\mathrm{AWGN}(R);\SNR)
\end{align*}
Moreover,
\begin{equation*}
 f_\mathrm{bnd}(d_\mathrm{typ}(R),\theta^r_\mathrm{AWGN}(R);\SNR)
\expeq e^{-n \,  E_\mathrm{AWGN}(R;\SNR) }.
\end{equation*}
\end{theorem}
Taking $\phi(R) = \theta_\mathrm{AWGN}(R;\SNR)$ in the preceding
theorem yields our final characterization of the error exponent of the
AWGN channel.

\begin{corollary}
\label{thrm:awgn_char}
Consider the sequence of ensembles of random spherical codes
$\{\Omega^{(n)}_\mathrm{III}(R)\}$.
If $\cR = \Rcone(\theta_\mathrm{AWGN}(R))$, then
Gallager's bounding technique \eqref{eqn:gallager_region}
for the average probability of error
satisfies the following two properties for $0 \le R \le C$:
\begin{equation*}
   \begin{aligned}
      1'.  \qquad &   \Prmrb \exple
                     \exp\left(-n \,  E_\mathrm{AWGN}(R;\SNR)\right)      \\
      2'.  \qquad &  \Prmub^r \exple
                    \Prmr        \\
   \end{aligned}
\end{equation*}
Further, if  $\cR = \Rcone(\theta_\mathrm{AWGN}(R))$ then
the typical error events occur with codewords at a distance
$d_\mathrm{typ}(R)$.
\end{corollary}

As in the case of the sphere-packing bound, one may ask whether there
are smaller regions than the cone $\Rcone(\theta_\mathrm{AWGN})$
that may be used to derive the same exponential upper bound on
$\Prmub^r$.   We extend our previous definition
of ``valid'' regions [cf.~\eqref{eqn:valid_sphere_def}] to be the
regions $\cR$ such that:
\begin{align}
&1.  \quad\pr\left(\bc + \bz \notin \cR \right) \expeq
\pr\left(\bc + \bz \notin \Rcone(\theta_\mathrm{AWGN}) \right) \\
&2.  \quad\cR \subset \Rcone(\theta_\mathrm{AWGN})
\end{align}
Recall that in order to show that the mod-$\Lambda$ channel can
achieve the sphere-packing error exponent for the AWGN channel we took
a scaling $\alpha$ that corresponded to a valid sphere.   More
precisely, in Section \ref{sec:sphere_region} we showed that one may
use the sphere tangent to the cone $\Rcone(\theta(R))$ and
achieve the same exponential upper bound on $\Prmr$.
Replacing $\theta(R)$ with $\theta_\mathrm{AWGN}(R)$ in that context one
may do the same.   However, in the sequel we show that the spherical
region $\Rcone(\theta_\mathrm{AWGN}(R))$ \emph{does not
  correspond to the optimal scaling} in the mod-$\Lambda$
channel.  Indeed, as we have seen for rates less than $\Rcrit$
the upper bound on the union bound, $f_\mathrm{bnd}$, dominates our bound
on the error exponent.   Hence, one would expect that the optimal
scaling would relate to the half angle $\thetaD{d}/2$ and not to the
half angle of the cone $\Rcone(\theta)$.   We show in the
sequel that the optimal scaling does indeed relate to the half angle
$\thetaD{d}/2$ and, for the ensemble $\Omega^{(n)}_\mathrm{I}(R)$, is
equal to
\begin{equation*}
   \alpha_\mathrm{AWGN}^r(R) = \asphere^*(\max\{R,\Rcrit\})
\end{equation*}
and, for the ensemble $\Omega^{(n)}_\mathrm{III}(R)$, is equal to
$\alpha_\mathrm{AWGN}(R;\SNR)$ defined in \eqref{def:alpha_awgn} at
the bottom of the page.\footnote{Recall $\alpha^*_s(R)$ was defined in
\eqref{eqn:alpha_star}, $\bcone^*(\theta,\Theta;\SNR)$ was defined in
\eqref{eqn:beta_star_d}, $\thetaD{d}$ was defined in
\eqref{eqn:thetad_def} and $\asphere(\beta,\theta)$ was defined in
\eqref{eqn:alpha_rel}.}
\begin{figure*}[b]
\hrulefill
\begin{equation}
   \alpha_\mathrm{AWGN}(R;\SNR)
    =
    \begin{cases}
       \asphere( \bcone^*(\theta(R),\thetaD{d_\mathrm{typ}(R)}/2;\SNR),
               \theta(R))
          & \text{ if $\Rcrit < R < C$,} \\
       \asphere( \SNR \cdot
          \bcone^*(\theta(R),\thetaD{d_\mathrm{typ}(R)}/2;\SNR),
               \thetaD{d_\mathrm{typ}(R)/2})
              & \text{ if $0 < R \le \Rcrit$.}
    \end{cases}
\label{def:alpha_awgn}
\end{equation}
\end{figure*}

We show in Appendix \ref{sec:sphere_details}
that the ensemble $\Omega^{(n)}_\mathrm{III}(R)$ achieves the full
AWGN error exponent $E_\mathrm{AWGN}(R;\SNR)$ using the scaling
$\alpha_\mathrm{AWGN}(R;\SNR)$.  We now provide our final
characterization of the error exponent of the mod-$\Lambda$ channel
and show that in general a scaling different than
$\alpha_\mathrm{AWGN}(R;\SNR)$ is needed to achieve our best bound on
the error exponent of the mod-$\Lambda$ channel.

\section{Error Exponents in the Mod-$\Lambda$ Channel}
      \label{sec:lattice}

In Section \ref{sec:lattice_def} we provided a simple proof that the
mod-$\Lambda$ channel achieves the sphere-packing error exponent {for
rates greater that the critical rate} under the assumption that
$\Prmb^{\lambda,r}_\mathrm{union} \le \Prmr^\lambda$.  Here we show
that this is indeed true and provide an exponential bound for
$\Prmb^{\lambda,r}_\mathrm{union}$ that is exponentially
equal to the random coding exponent $E^r_\mathrm{AWGN}(R;\SNR)$.  As
in Section \ref{sec:lattice_def} we provide a simple relation to the
derivation of the AWGN error exponent using spherical regions.  For
this reason we take the region $\cR$ to be a sphere of radius $\sqrt{n
P} \, r$ centered at the codeword.  We denote this region
$\cR_\lambda(r)$.

Note that $\cR_\lambda(r)$ is not an actual decoding region.
However, in order to choose the radius of the sphere $\cR_\lambda(r)$ we may use the same intuition that led to our choice
of the cone $\Rcone(\theta(R))$ in our derivation for the AWGN
channel.  That is, we can choose $r$ such that $\cR_\lambda(r)$ has a volume equal to the average volume of the
Voronoi under ML decoding.  Thus, for rates greater than critical rate
we consider $r = r^\alpha(R) = {\sin \theta(R)}/{\alpha} =
{\exp(-R)}/{\alpha}$.  Analogous to our definition of $R(\theta)$ we
let
\begin{equation*}
   R^\alpha(r) = - \log( \alpha \cdot r )
\end{equation*}
so that  $R^\alpha(r^\alpha(R) ) = R$.

We consider the ensemble of {random coset codes} that are drawn {\it
  i.i.d} {from a uniform distribution over the} Voronoi region of a
lattice $\Lambda$ that is good for quantization and good for
covering.  Recall from \eqref{eqn:average_region_lattice} that for any
code $\cC$ and any given codeword $\bc \in \cC$ we
have
\begin{equation*}
   \Prmb_e^\lambda
   \le \Prmr^\lambda + \Prmub^{\lambda,r}.
\end{equation*}
where
\begin{equation} \label{eqn:union_lattice}
  \Prmub^{\lambda,r} =  \E \,
  \frac{1}{|\cC|} \sum_{\bc \in \cC}
 \sum_{
                         \bc_e^\lambda \in \cC^{\lambda,r} \setminus
                         \bc
                           } \pr\left(
\cE_\alpha(\tilde{\by},\bc_e^\lambda)
             , \frac{\bc}{\alpha} + \bz'_\mathrm{eff} \in {\cR_\lambda(r)}
                 \right)
\end{equation}
and the expectation is taken over the ensemble
$\hat{\Omega}^{(\Lambda,n)}_0$ of {random coset} codes.  It is
important to note that for a fixed code each codeword, say $\bc$, and
its translates $\bc + \lambda$ for $\lambda \in \Lambda$ are
dependent.  However, by averaging over the ensemble of codes the
distribution of the codewords $\bc_e^\lambda$ are \emph{uniform over
  $\RR^n$} with a density\footnote{Note that in the scaled lattice
  $\Lambda/\alpha$ the codewords are uniform over $\RR^n$ with a
  density of $\alpha e^{n R}/\sqrt{n P}$} of $e^{n R}/\sqrt{n P}$.

\begin{figure*}
\begin{center}
   \begin{tabular}{cc}
      \begin{psfrags}
      \psfrag{M}[bl]{$\bb$}
      \psfrag{ME}[rb]{$\bc_e$}
      \psfrag{X}[cc]{$x$}
      \psfrag{X}[cc]{$x_l$}
      \psfrag{Y}[cc]{$y_l$}
      \psfrag{B}[rc]{$\beta \sqrt{nP}$}
      \psfrag{OA}[tc]{\small $\frac{1-\alpha}{\alpha}$}
      \psfrag{ER}[rc]{\small $r \sqrt{nP}$}
      \psfrag{TD}[lc]{\small $\Theta_p$}
      \psfrag{L}[cc]{\small $l \sqrt{nP}$}
      \psfrag{S}[cc]{\Large $\circledcirc$}
      \includegraphics[scale=0.261]{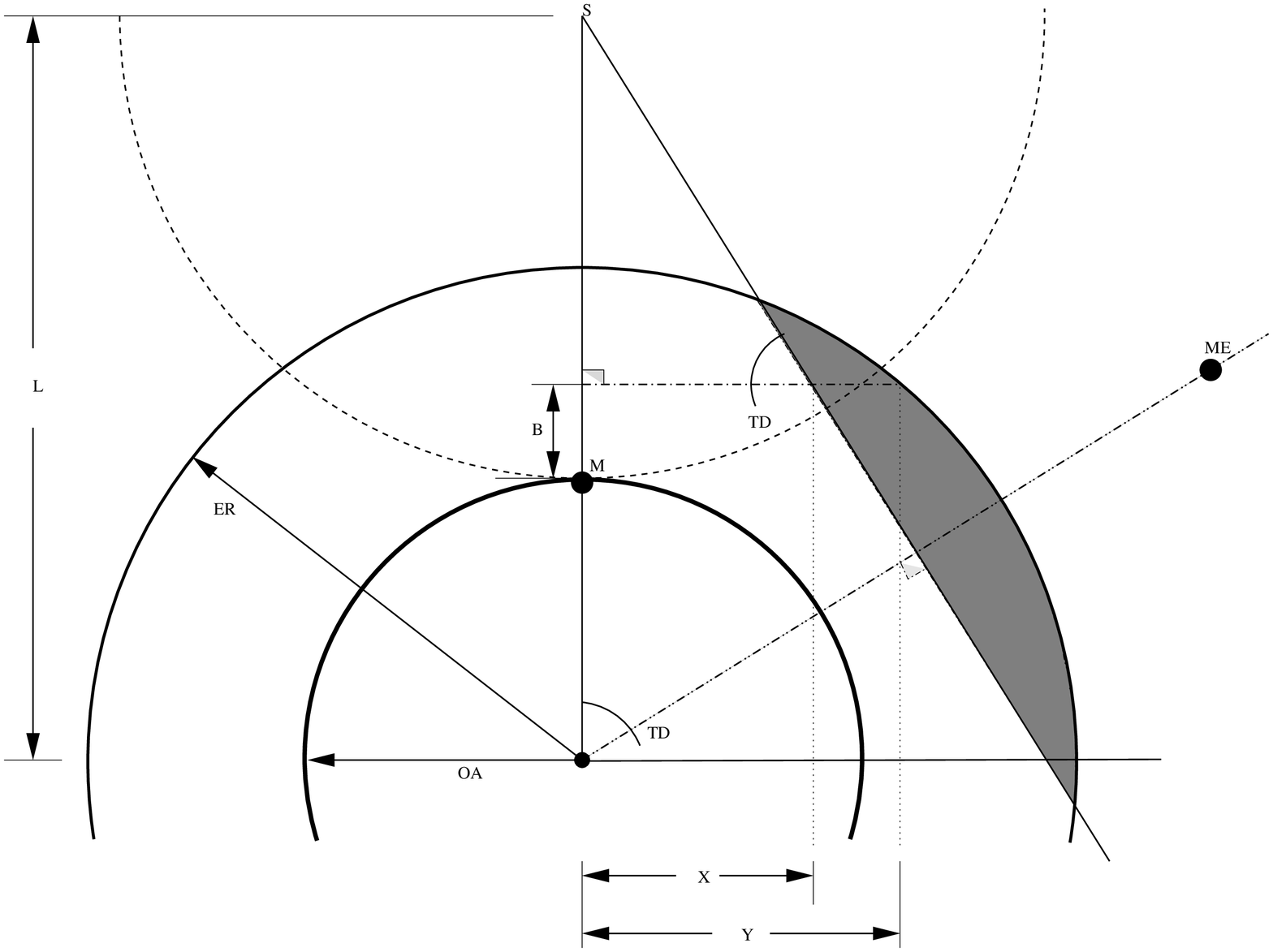}
      \end{psfrags}
&
      \begin{psfrags}
      \psfrag{M}[tc]{$\bm$}
      \psfrag{ME}[rb]{$\bm_e$}
      \psfrag{X}[cc]{$x$}
      \psfrag{X}[cc]{$\frac{d}{2}$}
      \psfrag{Y}[cc]{$y$}
      \psfrag{B}[cc]{$\beta$}
      \psfrag{OA}[tc]{$1-alpha$}
      \psfrag{ER}[lc]{$e^{-R}$}
      \psfrag{D}[lc]{$\cD_\lambda(r,d,l,\beta)$}
      \psfrag{O}[cc]{\Large $\circledcirc$}
      \includegraphics[scale=1.72]{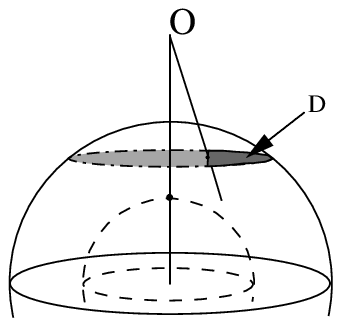}
      \end{psfrags}
   \end{tabular}
   \caption{The parameters for the derivation of the mod-$\Lambda$
              error exponent.  A relation to the AWGN error exponent can
              be made by thinking of $\circledcirc$ as the origin of
              a spherical code; see Fig.~\ref{fig:relation_fig}.    }
   \label{fig:big_picture_mlan}
\end{center}
\end{figure*}

Although the distribution of codewords are rotationally symmetric
since they are uniformly distributed over $\RR^n$, the dither
introduces an asymmetry for error events between codewords at a given
distance from a selected codeword.  It is necessary to consider
the orientation of any codeword to the dither when examining the error
events.   As depicted in Fig.~\ref{fig:big_picture_mlan}, let
$\Theta_p$ be the angle made between the direction of the dither and
the line from the original codeword to an arbitrary codeword at
distance $d$.   Note that in the case of a spherical code using the
relation shown in Fig.~\ref{fig:relation_fig}, $\Theta_p =
\frac{\pi}{2} - \thetaD{d}$ was always such that $\sin \Theta_p =
1-d^2/2$ since the codewords were constrained to lay on the surface of
a sphere [cf.~\eqref{eqn:thetad_def}].  However, in the case of the
mod-$\Lambda$ channel $\Theta_p$ varies independently of $d$.

We use the same method as used previously to further bound
\eqref{eqn:union_lattice}.  That is, we use the tangential sphere bound
and integrate with respect to the radial component of the noise.   As
before, we consider the region that is the intersection of
$\cR_\lambda(r)$ and the half space that orthogonally bisects the line
connecting the original codeword to any codeword at a distance $d$
\emph{and} angle $\Theta_p$.   As Fig.~\ref{fig:big_picture_mlan}
depicts, it is much simpler to consider this region when parametrized
by $l = \frac{d}{2 \cos \Theta_p}$, and thus we denote this region by
$\cD_\lambda(r,d,l)$.   Note that in order to find the dominating event
it is sufficient to optimize over the distance $d$ and $l$ since this
pair uniquely specifies $\Theta_p$.

It is shown in Appendix \ref{append:dist_dist_lattice} that
\begin{align}
   &\Prmub^{\lambda,r}
    \exple
   \mathop{ \max_{ (d,l) } }_{  0 \le d \le 2 r, l \ge K_\alpha }
   \pr\left( \frac\bc{\alpha} + \bz_\mathrm{eff}'
            \in \cD_\lambda(r,d,l) \right)  \notag \\
    & \qquad \qquad \times \exp\left( n R + (n-1) \log
    \left( \alpha d \sqrt{1-\frac{d^2}{4 l^2}} \right) \right).
   \label{eqn:mlan_bound}
\end{align}
In order to arrive at an exponential bound to the
right-hand side of \eqref{eqn:mlan_bound} we begin
by providing an exponential bound for
$\pr\left(  \bc/{\alpha} + \bz_\mathrm{eff}' \in
\cD_\lambda(r,d,l) \, | \, \bu \right)$.  We use the
tangential sphere bound as in Section \ref{sec:low_rates} by defining the
radial
direction to be the direction of the dither $\bu$.  It is clear due to the
rotational symmetry that
\begin{equation*}
   \pr\left(  \bc/{\alpha} + \bz_\mathrm{eff}' \in
\cD_\lambda(r,d,l) \, | \, \bu \right) =
\pr\left(  \bc/{\alpha} + \bz_\mathrm{eff}' \in
\cD_\lambda(r,d,l)\right)
\end{equation*}
and henceforth we write $\pr\left( \bc/{\alpha} + \bz_\mathrm{eff}'
\in \cD_\lambda(r,d,l)\right)$.   From Fig.~\ref{fig:big_picture_mlan}
we see that conditioned on the event that the radial component of the
noise has a magnitude of $\sqrt{n P} \cdot \beta_\lambda$, we have
that $ \bc/{\alpha} + \bz_\mathrm{eff}' \in \cD_\lambda(r,d,l)$ if the
second component is greater than $x_\lambda$ while the sum of the
second through $n$th component is less than $y_\lambda$ where
\begin{equation*}
   x_\lambda = x_\lambda(\beta_\lambda,l,\alpha)
       =  \frac{ l - (\beta_\lambda + K_\alpha ) }{
            \sqrt{ \frac{4 l^2}{d^2} -1 }
         }
\end{equation*}
and
\begin{equation*}
   y_\lambda^2 = y_\lambda^2(\beta_\lambda,r,\alpha)
         =  r^2 -  \left( \beta_\lambda+ K_\alpha \right)^2.
\end{equation*}
Thus, applying Proposition \ref{prop:chernoff_constrain}
and Proposition \ref{prop:voron_ball} we have
as in \eqref{eqn:cone_exponent}
\begin{align}
& \pr\left(  \frac\bc{\alpha} + \bz_\mathrm{eff}'
\in \cD_\lambda(r,d,l) \right) \notag \\
 & \qquad = e^{o(1)} \pr\left(   \frac\bc{\alpha} + \bz_\mathrm{eff}''
       \in \cD_\lambda(r,d,l) \right) \\
 & \qquad \exple \max_{  \beta_\lambda }
          \exp\bigl( -n \,
          {\Ebd}\left(\beta_\lambda,x_\lambda, y_\lambda);
                     \SNR \right) \bigr)
 \label{eqn:mlan_exponent}
\end{align}
Thus, \eqref{eqn:mlan_bound} becomes
\begin{equation}
   \Prmub^{\lambda,r}
    \exple
   \max_{ (d,l) }
   \max_{  \beta_\lambda }  \,
   \exp\bigl( -n \,
    E_\mathrm{bnd}^\lambda(r,K_\alpha,l,d,\beta,R;\SNR) \bigr) .
   \label{eqn:mlan_bound_final}
\end{equation}
where $ E_\mathrm{bnd}^\lambda$ is defined in \eqref{eqn:Ecirc_def} at the
bottom of the page.
\begin{figure*}[b]
\hrulefill
\begin{equation} \label{eqn:Ecirc_def}
   E_\mathrm{bnd}^\lambda(r,K_\alpha,l,d,\beta;\SNR)
  = \Ebd(\beta,x_\lambda,y_\lambda;\SNR) - \frac{1}{2}  \log{\left[
            \frac{d^2}{l^2} \left(1 - \frac{d^2}{4 l^2}\right)
                        \right]}
    - \frac{1}{2} \log{ \left[ \frac{l^2}{ (1 + K_\alpha)^2} \right] } - R.
\end{equation}
\end{figure*}
Then, it is simple to check that the $\beta_\lambda$ that maximizes
\eqref{eqn:mlan_bound_final} for a fixed $d$ and $l$ is
\begin{equation}
 \beta_\lambda^*(r,d,l;\SNR)
  = \left\{
                \begin{array}{lc}
 \beta_\circ^*(r,d,l ; \SNR ) &
\text{if }  R^\alpha(r) > R^\alpha_\mathrm{cr}(d,l)
\\
                    {d^2}/{4 l^2} \cdot \left( l - K_\alpha \right)
                      & \text{ otherwise }
                \end{array}
                \right.
\end{equation}
where $\beta_\circ^*(r,d,l ; \SNR )$ satisfies \eqref{eqn:key_beta} at
the bottom of the page 
\begin{figure*}[b]
\hrulefill
\begin{equation}
\label{eqn:key_beta}
     l - K_\alpha - \beta_\circ^*(r,d,l ; \SNR )
     = l\left( 1 -   \frac{ d^2 }{ 4 l^2 } \right)
        + \frac{1}{ 2 K_\alpha \SNR} -
        \sqrt{\frac{1}{4 K_\alpha^{2}  \, \SNR^2}
              + \left( r^2 - \frac{d^2}{4} \right)
                \left(1 - \frac{d^2}{4 l^2} \right) }.
\end{equation}
\end{figure*}
and in turn where
\begin{equation*}
    R^\alpha_\mathrm{cr}(d,l) \myDef - \frac{1}{2}
           \log\left[ \frac{1}{1+K_\alpha}\left( \frac{d^2}{4} +
                       K_\alpha^2 \left( 1 -  \frac{d^2}{4 l^2 }  \right) +
                       \frac{1}{\SNR} \right)
               \right].
\end{equation*}

In our derivation of an exponential bound for $\Prmr$ we
were able to show that the typical error events in the AWGN channel
and the mod-$\Lambda$ channel \emph{coincide} if we use the scaling
that is equivalent to a valid sphere.   We now show that this is again
the case for $\Prmub^{\lambda,r}$ and thus the
mod-$\Lambda$ channel can achieve the random coding error exponent,
$E^r_\mathrm{AWGN}(R;\SNR)$.
We first consider what parameters one must choose in order for the
geometry of Fig.~\ref{fig:big_picture_mlan} to equal that in
Fig.~\ref{fig:big_picture_awgn} as done in Fig.~\ref{fig:relation_fig}.
Examining these figures it is clear that for this to be true
we must have that
\begin{equation} \label{eqn:key_equal}
 l - K_\alpha - \beta_\circ^*(r,d_\lambda^*, l_{\lambda}^*;\SNR) ; \SNR )
   = 1 + \bcone^*(\theta; \SNR)
\end{equation}
where $d_\lambda^* = d_\lambda^*(r,l,\alpha;\SNR)$
and $l_{\lambda}^* = l_{\lambda}^*(r,\alpha;\SNR)$ are the
distance and value of $l$ that maximize
$\Prmub^{\lambda,r}$
respectively.
Further, it is clear that if
$r = (\sin \theta)/\alpha$
then one must have
$\Theta_p = \frac{\pi}{2} - \frac{\thetaD{d}}{2}$ and $l = 1/\alpha$ for the
geometry to agree.
Thus, to show that mod-$\Lambda$ channel is able
to achieve the random coding error exponent using our relation
we would require that
\begin{equation*}
     l_{\lambda}^*\left( \frac{\sin \theta^r_\mathrm{AWGN }(R) }{
         \alpha^r_\mathrm{AWGN}(R) } ,\alpha^r_\mathrm{AWGN}(R);
       \SNR \right) = \frac{1}{\alpha^r_\mathrm{AWGN}(R) }
\end{equation*}
and $d_\lambda^*$ to satisfy \eqref{eqn:lattice_dist} at the bottom of the
page
\begin{figure*}[b]
\hrulefill
\begin{equation} \label{eqn:lattice_dist}
   \alpha^r_\mathrm{AWGN}(R)
  \cdot
  d_\lambda^*\left( \frac{\sin \theta^r_\mathrm{AWGN }(R) }{
         \alpha^r_\mathrm{AWGN}(R) } ,
        \frac{1}{\alpha^r_\mathrm{AWGN}(R) }
       ,\alpha^r_\mathrm{AWGN}(R);
       \SNR \right)
  = \left\{
    \begin{array}{lc}
       \sqrt{2} \sin \theta^r_\mathrm{AWGN}  & \text{if }
         R > \Rcrit \\
        d_\mathrm{crit}
                      & \text{ otherwise }
                \end{array}
                \right.
\end{equation}
\end{figure*}

and further that these values satisfy \eqref{eqn:key_equal} for
$\theta = \theta^r_\mathrm{AWGN}$.  It is easy to check that these
values do indeed satisfy \eqref{eqn:key_equal} through direct
substitution.  We provide the general derivation of the parameters $
l_{\lambda}^*$ and $ d_\lambda^*$ for general $\alpha$ in Appendix
\ref{sec:lattice_details}.  The case when $\alpha =
\alpha^r_\mathrm{AWGN}(R)$ is easy to verify and leads to the
following theorem.

\begin{theorem}
\label{thrm:mlan_char}
Consider the sequence of random {random coset} ensembles $\{
\hat{\Omega}^{(\Lambda_n,n)}_0 \}$ where $\{\Lambda_n\}$ is a sequence
of lattices that are good for covering and quantization.  If $\cR =
\cR_\lambda(r^\alpha(R))$ and $\alpha = \alpha^r_\mathrm{AWGN}(R)$,
then Gallager's bounding technique \eqref{eqn:gallager_region} for the
average probability of error satisfies the following two properties
for $0 \le R \le C$:
\begin{equation*}
   \begin{aligned}
      1'.  \qquad &   \Prmub^{\lambda,r}
        \exple
                     \exp\left(-n \,  E^r_\mathrm{AWGN}(R;SNR)\right)        \\
      2'.  \qquad &  \Prmub^{\lambda,r}
        \expge \Prmr        \\
   \end{aligned}
\end{equation*}
where $E^r_\mathrm{AWGN}(R)$ was defined in \eqref{eqn:random_exp}.
\end{theorem}
A proof is provided in Appendix \ref{sec:lattice_details}.

Recall that when we considered the AWGN error exponent, the random
coding error exponent $E_\mathrm{AWGN}^r(R;\SNR)$ could be improved for
some rates by considering an ensemble of codes with minimum distance,
$d_\mathrm{min}(R)$.  It is natural to expect that the same can be done
here and indeed it can.   That is, as done in Section
\ref{sec:low_rates}, we can attempt to improve on the random coding
error exponent by considering the ensembles of rate $R$ {coset} codes
that meet a constraint on the minimum distance, which we denote as
$r_\Omega(R)$.   That is, the ensembles
$\hat{\Omega}^{(\Lambda,n)}_r(R) = ( \hat{\Omega}^{(\Lambda,n)},
r_\Omega(R) )$ (provided such an ensemble exists).  Note, for the
mod-$\Lambda$ channel we must consider the minimum distance for codes
distributed over $\RR^N$ rather than over the unit sphere.

In order to determine which $r_\Omega(R)$ led to valid ensembles
it is often easier to study a normalized version of the minimum distance.
Define, $\rho^{ }_\Lambda$, as
\begin{equation*}
 \rho^{ }_\Lambda = \frac{ r_\Omega(R) }{r_\Lambda^\mathrm{eff}},
\end{equation*}
where $r_\Lambda^\mathrm{eff}$ was defined in \eqref{eqn:reff_def}.
It is a classic problem to determine the largest possible
value for $\rho^{ }_\Lambda$.   In this direction, let
\begin{equation*}
   \rho^{ } = \limsup_{n \to \infty} \, \sup_{\Lambda}
   \rho^{ }_\Lambda.
\end{equation*}
Then, we can improve upon the random coding exponent for rates such
that $\alpha \, {d_\lambda^*} < 2 \rho^{ } { {r_\Lambda^\mathrm{eff}}}/{
  \sqrt{n P} }$.   From \eqref{eqn:lattice_dist} if $\alpha =
\alpha^r_\mathrm{AWGN}(R)$ this is equivalent to
\begin{equation} \label{eqn:rmin_eq}
  d_\mathrm{crit} \le 2 \rho^{ } \frac{{r_\Lambda^\mathrm{eff}}}{\sqrt{n P}} \\
                \le 2 \rho^{ } \exp(-R)
\end{equation}
if $\rho^{ } < 1/\sqrt{2}$ and
where the last inequality is satisfied with equality if $\Lambda$ is good
for covering.
To date the best known bounds on $\rho^{ }$ are \cite{ErezGoodLattice}
\begin{equation*}
   \frac{1}{2} \le \rho^{ } \le 0.660211...
\end{equation*}
Henceforth we consider ensembles of codes such that $r_\Omega(R) \le
\exp( - R )$.  In particular, we consider the ensemble of {coset} codes
\begin{equation*}
   \hat{\Omega}^{(\Lambda,n)}_\mathrm{II}(R) =
   (\hat{\Omega}^{(\Lambda,n)}_{0}, e^{-R})
\end{equation*}
where $\Lambda$ is good for covering and quantization.

We now return to our original problem of improving upon the random coding
error exponent.  Begin by noting that for the ensemble
$\hat{\Omega}^{(\Lambda,n)}_\mathrm{II}(R)$ if $\Lambda$ is good for covering
we can improve upon the random coding exponent for rates
$R < R_\mathrm{II}$ where
\begin{equation*}
R_\mathrm{II} = \max\left\{ 0, -
                    \log\left( {d_\mathrm{crit}} \right)
                 \right\}.
\end{equation*}
Further, note the definition of
$\hat{\Omega}^{(\Lambda,n)}_\mathrm{II}(R)$ is similar to the
definition of ${\Omega}^{(n)}_\mathrm{II}(R)$ in \eqref{eqn:e_II} in
Section \ref{sec:low_rates}.  In fact, one can use similar arguments
to those leading to Theorem \ref{thrm:mlan_char} to show that the
typical error events of $\hat{\Omega}^{(\Lambda,n)}_\mathrm{II}(R)$
and ${\Omega}^{(n)}_\mathrm{II}(R)$ coincide given the appropriate
scaling for the lattice.  In this direction we provide the following
the following definition.  Let, for any $r_\Omega(R) \le \exp( -R)$,
\begin{equation*}
   K_\alpha^* = K_\alpha^*(r_\Omega(R);\SNR)
\end{equation*}
where $K_\alpha^*(r_\Omega(R);\SNR)$ is defined in \eqref{def:Kalpha_opt}
at the bottom of the page,
\begin{figure*}[b]
\hrulefill
\begin{equation} \label{def:Kalpha_opt}
   K_\alpha^*(r_\Omega(R);\SNR) =
   \begin{cases}
      {({1-\alpha^*_s(\theta(R))})}/{{\alpha^*_s(\theta(R))}}
        & \text{if $\Rcrit(\theta) < R < C$} \\
      {((1-d_\mathrm{crit}^2/4) \, \SNR)^{-1}} & \text{if $R \le \Rcrit$ and $d_\mathrm{crit} > d_\Omega(R)$} \\
      ((1-r_{\Omega}(R)^2/4) \, \SNR)^{-1} & \text{if $R \le \Rcrit$ and $d_\mathrm{crit} \le d_\Omega(R)$}
   \end{cases}
\end{equation}
\end{figure*}
and let
\begin{equation*}
   \alpha_\lambda^*(r_{\Omega}(R);\SNR)
   = \frac{1}{1 +  K_\alpha^*(r_\Omega(R);\SNR))}.
\end{equation*}
In Appendix \ref{sec:lattice_details}, we show that choosing
$\alpha = \alpha_\lambda^*(r_{\Omega}(R);\SNR)$ one has
\begin{equation} \label{eqn:lambda_key}
   l_\lambda^*( r^\alpha(R), \alpha_\lambda^*(r_{\Omega}(R);\SNR); \SNR )
   = 1 + K_\alpha^*(r_\Omega(R);\SNR)
\end{equation}
yielding
\begin{align}
   &\mathop{ \min_{d \ge r_\Omega(R)} }_{ l \ge K_\alpha^* }
   \min_{ \beta \ge -1 }
    E_\mathrm{bnd}^\lambda(  (1+K_\alpha^*) \cdot \sin \theta, K_\alpha^*,
   \beta,R;\SNR)  \notag \\
   &\qquad \quad = \min_{ d \ge r_\Omega(R)}  \min_{ \beta \ge -1 }
    E_\mathrm{bnd}(\theta,d,\beta,R;\SNR) \label{eqn:opt_awgn} \\
   &\qquad \quad = \Et_\mathrm{bnd}( r_\Omega(R),\theta;\SNR)
   \label{eqn:opt_mlan}
\end{align}
We note that \eqref{eqn:lambda_key} -- \eqref{eqn:opt_mlan} is the formal
statement of our geometric equivalence depicted in
Fig.~\ref{fig:relation_fig}.   That is, choosing the appropriate
scaling, $\alpha_\lambda^*(r_{\Omega}(R);\SNR)$, the typical
error events in the mod-$\Lambda$ and AWGN channels coincide.  This yields the
following theorem.

\begin{theorem}
Consider the sequence of ensembles of random lattice codes
$(\hat{\Omega}^{(\Lambda,n)},r_\Omega(R))$ where $\{\Lambda_n\}$ is a
sequence of lattices that are good for covering and quantization.   If
$\cR = \cR_\lambda(r^\alpha(R))$, $0 \le R < C$, and
$\alpha = \alpha_\lambda^*(r_{\Omega}(R);\SNR)$, then Gallager's
bounding technique \eqref{eqn:gallager_region} for the average
probability of error is exponentially equal to that of the ensemble
$(\hat{\Omega}^{(n)},r_\Omega(R))$.   That is,
\begin{equation*}
   \Prmub^{\lambda,r} \le
     f_\mathrm{bnd}(r_\Omega(R),\theta(R);\SNR)
\end{equation*}
with $f_\mathrm{bnd}(d,\theta;\SNR)$ as defined in \eqref{eqn:fbdn}.  Moreover,
with $\alpha= \alpha_\lambda^*(r_{\Omega}(R);\SNR)$,
$l_\lambda^*(r,\alpha;\SNR) = 1/\alpha$ and the typical error
events for the ensemble $(\hat{\Omega}^{(\Lambda,n)},r_\Omega(R))$ and
$(\hat{\Omega}^{(n)},r_\Omega(R))$ coincide.
\end{theorem}
A proof is provided in Appendix \ref{sec:lattice_details}.

In order to provide the desired relation to the ensemble
${\Omega}^{(n)}_\mathrm{II}(R)$
we let
\begin{equation*}
\alpha_\Lambda(R;\SNR) = \frac{1}{1+K_\alpha^*(\exp(-R);\SNR)}.
\end{equation*}

\begin{corollary}
\label{thrm:mlan_char_fin}
Consider the sequence of random {coset} ensembles $\{
\hat{\Omega}^{(\Lambda_n,n)}_\mathrm{II} \}$ where $\{\Lambda_n\}$ is a
sequence of lattices that are good for covering and quantization.  If
$\cR = \cR_\lambda(r^\alpha(R))$ and $\alpha =
\alpha_{\Lambda}(R;\SNR)$, then Gallager's bounding technique
\eqref{eqn:gallager_region} for the average probability of error of
the ensemble $\hat{\Omega}^{(\Lambda_n,n)}_\mathrm{II}$ is exponentially
equal to that of the ensemble ${\Omega}^{(n)}_\mathrm{II}(R)$.
Moreover,
\begin{equation*}
 \begin{aligned}
 \exp( -n E^r_\mathrm{AWGN}(R;\SNR) )
 &\expeq   \exp( -n E_\mathrm{I}(R;\SNR)    ) \\
 &\exple \exp( -n E_\mathrm{II}(R;\SNR)   ) \\
 &\expeq   \exp( -n E_{\Lambda}(R;\SNR)  ) \\
 &\exple \exp( -n E_\mathrm{III}(R;\SNR)  ) \\
 &\expeq   \exp( -n E_\mathrm{AWGN}(R;\SNR) )
 \end{aligned}
\end{equation*}
\end{corollary}

As done in Section \ref{sec:low_rates} it is natural to ask whether
one can improve upon the error exponent for the mod-$\Lambda$ with a
different choice of region than that taken in Corollary
\ref{thrm:mlan_char_fin} as $\Prmub^{\lambda,r} \expge \Prmr$.   We now
characterize the freedom one has in this choice.   Let
\begin{equation*}
d^\mathrm{II}_\mathrm{typ}(R) \myDef \left\{
        \begin{aligned}
              \exp(-R) & \qquad \text{ if } \, 0 \le R \le R_\mathrm{II} \\
              d_\mathrm{crit}  &  \qquad \text{ if } \, R_\mathrm{II} < R \le \Rcrit \\
           \sqrt{2} \exp(-R) &  \qquad \text{ if } \, \Rcrit < R  \le C \\
        \end{aligned}
        \right.
\end{equation*}
and let $\theta_\Lambda(R;\SNR)$ be defined as in \eqref{def:theta_mlan} at
the bottom of the page.
\begin{figure*}[b]
\hrulefill
\begin{equation} \label{def:theta_mlan}
   \theta_\Lambda(R;\SNR) =
   \left\{
   \begin{aligned}
      \theta_\zeta(K_\zeta(d^\mathrm{II}_\mathrm{typ}(R);R,\SNR);\SNR) &
      \qquad \text{ if } 0 \le R < \Rcrit \\
      \arcsin \exp\left( - R \right) &
      \qquad \text{ if } \Rcrit \le R \le C
   \end{aligned}
   \right.
\end{equation}
\end{figure*}
In turn, let
\begin{equation*}
   r_\Lambda^\alpha(R;\SNR) = \frac{\exp( - \sin \theta_\Lambda(R;\SNR))}{
                                  \alpha_\Lambda(R;\SNR)}
\end{equation*}
That is, $r_\Lambda^\alpha(R;\SNR)$ is the smallest radius such that
$f_\mathrm{bnd}(d^\mathrm{II}_\mathrm{typ}(R);R,\SNR) \expge
\Prmr$.  This is characterized in the following
theorem.

\begin{theorem}
\label{thrm:mlan_char_total}
Consider the sequence of random {random coset} ensembles $\{
\hat{\Omega}^{(\Lambda_n,n)}_0 \}$ where $\{\Lambda_n\}$ is a sequence
of lattices that are good for covering and quantization.  If $\cR =
\cR_\lambda(r^\alpha_\Lambda(R))$ and $\alpha =
\alpha_\Lambda(R;\SNR)$, then Gallager's bounding technique
\eqref{eqn:gallager_region} for the average probability of error
satisfies the following two properties for $0 \le R \le C$:
\begin{equation*}
   \begin{aligned}
      1'.  \qquad &   \Prmub^{\lambda,r}
        \exple
                     \exp\left(-n \,  E_\mathrm{II}(R;SNR)\right)        \\
      2'.  \qquad &  \Prmub^{\lambda,r}
        \exple \Prmr        \\
   \end{aligned}
\end{equation*}
\end{theorem}

It is easy to see by examining \eqref{eqn:rmin_eq} and
\eqref{eqn:dmin_eq} that using our derivation, the error exponent of
the mod-$\Lambda$ channel would be to equal that of the AWGN channel
had the minimum distance of the coset code been equal to that of the
  spherical code.   However, the best known lower bound on the minimum
distance in a constellation in $\RR^N$ with a given density is less
than that of a spherical code \cite{PoltyrevUncon}.   Thus, the error
exponent for the mod-$\Lambda$ channel cannot be shown to be
equivalent to that of the AWGN channel using this approach for rates
less than $R_x$.  In fact, by examining the exponents at low rates one
can see that even the best known upper bound for $\rho^{ }$ is not
sufficient to achieve the AWGN error exponent for all rates less than
$R_x$.

\section{Conclusion}

It remains an open problem to show whether the mod-$\Lambda$ channel
can achieve the expurgated error exponent for all rates.   Note, that
in our derivation of the error exponent we used the sub-optimal
Euclidean distance decoder.   One may ask whether the closest coset
decoder or a true ML decoder could achieve the expurgated error
exponent.   It is our conjecture that this in fact cannot be done.
This conjecture is motivated by the fact that it was shown in
\eqref{eqn:opt_awgn}--\eqref{eqn:opt_mlan} that using a sub-optimal
Euclidean distance decoder a linear scaling existed such that the
mod-$\Lambda$ channel meets the best known lower bound on the
reliability of the AWGN channel \emph{if} the mod-$\Lambda$ channel
and AWGN channel codes have the same minimum distance.   However, as it
is known that the minimum distance of a lattice is less than that of a
spherical code at low rates it is unlikely that the mod-$\Lambda$
channel can achieve the expurgated error exponent for all rates using
a ML decoder.   However, the mod-$\Lambda$ channel is itself suboptimal
in the fact that it uses a linear estimator at the receiver.   It may
be possible to show that lattice encoding and decoding could achieve
the expurgated error exponent by using a non-linear receiver.

\appendices

\section{Proof of \eqref{eqn:rel} }
\label{append:rel}

Begin by noting that
\begin{equation*}
  (1+\beta^*) = \frac{1 + \rho_G}{\SNR} (\beta_G - 1)
\end{equation*}
and
\begin{equation*}
 \left( \frac{(1+\beta^*)^2}{ \cos^2 \theta(R)}
             - (1+\beta^*) \right) = \frac{1}{\SNR}.
\end{equation*}
Then, we have
\begin{equation*}
\begin{aligned}
   1 + \rho_G =& \SNR e^{-2 R} \frac{1 + \beta^*}{1-e^{-2 R}}  \\
              =& \SNR \frac{e^{-2 R} }{1-e^{-2 R}} (1+\beta^*)  \\
              =& \frac{r^2(\beta^*) \SNR}{(1+\beta^*)}
\end{aligned}
\end{equation*}
so that
\begin{equation*}
\begin{aligned}
   \beta_G - \frac{\SNR}{1 + \rho_G}
              =&  e^{2 R} - \frac{(1+\beta^*)}{r^2(\beta^*)} \\
              =&  \frac{1}{r^2(\beta^*)}
             \left( e^{2 R} r^2(\beta^*) - (1+\beta^*) \right) \\
              =&  \frac{1}{r^2(\beta^*)}
             \left( \frac{(1+\beta^*)^2}{ \cos^2 \theta(R)}
             - (1+\beta^*) \right) \\
              =& \frac{1}{r^2(\beta^*) \SNR }.
\end{aligned}
\end{equation*}
Now,
\begin{equation*}
  \begin{aligned}
  &2 E_v( \beta^* \SNR ) \\ &= \SNR \, (\beta^*)^2 \\
                        &= \SNR \,(-1 + (1 +\beta^*) )^2 \\
                        &= \SNR \, (1 - 2 (1 +\beta^*) +
                                      (1 +\beta^*)^2 ) \\
                        &= \SNR \, \left( 1 - 2 (1 +\beta^*)
                                      +
           (1 +\beta^*)^2
         \left(\frac{1}{\cos^2 \theta(R)} - \tan^2 \theta(R) \right) \right) \\
                        &= \SNR \, \left( 1 - (1 +\beta^*)
                   + \frac{1}{\SNR} - r^2(\beta^*) \right) \\
                        &= \SNR - \SNR \, (1+\beta^*)
                                - 2 E_h( r^2(\beta^*) \SNR )
                                - \log( r^2(\beta^*) \SNR )
  \end{aligned}
\end{equation*}
Thus,
\begin{equation*}
  \begin{aligned}
    &2 E_v( \beta^* \SNR ) + 2 E_h( r^2(\beta^*) \SNR )  \\
    &= \SNR - \SNR \, (1+\beta^*) - \log( r^2(\beta^*) \SNR ) \\
    &= \SNR - (1-\beta_G)(1+\rho_G) + \log\left(
                                        \beta_G - \frac{\SNR}{1 + \rho_G}
                                     \right) \\
    &= 2 E_\mathrm{sp}(R;\SNR)
  \end{aligned}
\end{equation*}

\section{Derivation of Maximizing Parameters for Random Coding Exponent}
   \label{sec:cone} \label{append:line}

We now provide a derivation of the minimizing parameters for the
random coding exponent.   Begin by examining \eqref{eqn:expon_Ebd} and
note that the exponent $E_\mathrm{d}(\beta,x,y;\SNR)$ has two cases based
on the values of $x$ and $y$ relative to $\SNR$.   Note that by using
the inequality $\log x \ge x - 1$ for $x > 1$ we have that
\begin{equation*}
  {\SNR} \cdot ( \beta^2 + y^2 ) -
             \log ( e \SNR ( y^2 - x^2)) \ge
  {\SNR \cdot \left( x^2 + \beta^2 \right) }.
\end{equation*}
Thus, since $\Ebd$ is continuous and increasing in $\beta$ we can always
considering minimizing $\Ebd$ under the assumption $y^2 - x^2 \le 1/\SNR$ and
provide an improvement if this is not the case.   However, since we have
freedom in our choice of $\theta$ it should be clear that
for any $\bcone$ it is always advantages to pick $\theta$ such that
$y_c(\bcone,\thetaD{d})^2 - x_c(\bcone,\thetaD{d})^2 \ge 1/\SNR$.  We examine
how this may be done after first considering the minimization under the
assumption  $y^2 - x^2 \le 1/\SNR$.

Consider the minimization of the exponent $\Ebd(R)$ under the assumption
$y_c^2 - x_c^2 \le 1/\SNR$.  In this case
the exponent $\Ebd(R)$ is greatly simplified
and in fact simply reduces to the exponent of a one dimensional Gaussian.   That
is, in the case that $y_c^2 - x_c^2 \le 1/\SNR$ the pairwise
error probability between two codewords at a distance $d$,
say $\bc$ and $\bc_e$, is dominated by the probability that a
one dimensional component of the noise crosses the ML decoding plane and
stays in the cone $\Rcone(\theta)$.

Now, using the tangential
sphere bound we have that the minimal $\bcone$ is
\begin{equation*}
   1 + \bcone^* = 1 - \frac{ \tan^2 \frac{\thetaD{d}}{2} }{ 1 + \tan^2
                     \frac{\thetaD{d}}{2} }
             = 1 - \sin^2 \frac{\thetaD{d}}{2} = \cos^2 \frac{\thetaD{d}}{2}
\end{equation*}
and \eqref{eqn:cone_bound} becomes \eqref{eqn:line_d} at the bottom of the
page,
\begin{figure*}[b]
\hrulefill
\begin{equation}
   \Prmub^r
   \le \exp\left( -n \min_{ 0 \le d \le 2 }  \frac{\SNR}{2} \cdot \sin^2
          \frac{\thetaD{d}}{2}
        - \log\left( \sin \thetaD{d} \right) - R \right)
\label{eqn:line_d}
\end{equation}
\end{figure*}
which is what one would have if one considered the noise along  the
line from $\bc$ to $\bc_e$.
Thus, minimizing  \eqref{eqn:line_d} we find that
\begin{equation*}
   d_c^* =
   d_\mathrm{crit} = \sqrt{ 2 + \frac{4}{\SNR} - 2 \sqrt{1 + \frac{4}{\SNR^2}}}.
\end{equation*}
Note that  $d_c^*$ is independent of the choice of the half angle of the
cone $\Rcone(\theta)$.   Thus, by choosing $\theta$ properly we can
guarantee that
\begin{equation} \label{eqn:line_cond}
y_c(\bcone^*,\theta)^2
- x_c(\bcone^*,\theta)^2 \ge 1/\SNR.
\end{equation}
However, in order to optimize the overall probability of error we may
not take $\theta$ arbitrarily since we must have $\Prmb^r_\mathrm{union} = \Prmr$.   Thus, for some rates
\eqref{eqn:line_cond} may not hold.  Indeed, closely examining
\eqref{eqn:line_cond} for $\theta = \theta(R)$ we have that if
\eqref{eqn:line_cond} is true then
$$R > 1/2 \log\left( \frac{1}{2} + \frac{\SNR}{4} + \frac{1}{2}
\sqrt{1 + \frac{\SNR^2}{4}} \right) = \Rcrit.$$ Thus, if $R >
\Rcrit$ then $d_\mathrm{crit}$ is not the dominating error
event.  In this case we must consider optimizing $\Ebd(R)$ with $y_c^2
- x_c^2 \ge 1/\SNR$.

It is a simple computation to see that the minimizing $\bcone$ in the
exponent of \eqref{eqn:cone_exponent} in the case that $y_c^2 - x_c^2
\ge 1/\SNR$ is
\begin{equation}
   1 + \bcone^*(\theta,\thetaD{d}; \SNR ) = \frac{\cos^2 \theta}{2} +
       \frac{\cos \theta}{2}
      \sqrt{  \cos^2 \theta + \frac{4}{\SNR} }.
\end{equation}
Substituting $\bcone^*$ in to \eqref{eqn:cone_bound} we have
\eqref{eqn:union_sphere} at the bottom of the page.
\begin{figure*}
\hrulefill
\begin{equation} \label{eqn:union_sphere}
 \Prmub^r
    \le \exp\left( -n \min_{ 0 \le d \le 2 }
           {\Ebd}\left(\bcone^*,x_c, y_c ; \SNR \right)
    -\log \left( d \sqrt{1 - \frac{d^2}{4} } \right) - R \right)
\end{equation}
\end{figure*}
Hence, in the case that $y_c^2 - x_c^2 \ge 1/\SNR$, we have
the minimizing  $d$ as
\begin{equation}
   d_c^* = \frac{ \sqrt{2} y_c }{ \sqrt{ (1+\bcone^*)^2 + y_c^2 } }
         = \frac{ \sqrt{2} \tan \theta(R) }{ \sqrt{ 1 + \tan^2 \theta(R) } }
         = { \sqrt{2} \sin \theta }
\label{eqn:cone_packing_error}
\end{equation}

\section{Derivation of Parameters Minimizing the Mod-$\Lambda$ Error Exponent}
   \label{sec:lattice_details}

As done in Appendix \ref{append:line}, we consider the two cases of
$\beta_\lambda^*$ separately.   First, we consider the case where
$\beta_\lambda^*= \beta_\circ^*(r,d,l ; \SNR )$ and begin by
examining the partial derivative of
$ E_\mathrm{bnd}^\lambda(r,K_\alpha,l,d,\beta^*_\lambda;\SNR)$
with respect to $l$.   One can check that this partial is zero if
\begin{equation*}
   l = l_\circ^*(r,K_\alpha;\SNR) \, \textnormal{or } \,
   l = -l_\circ^*(r,K_\alpha;\SNR) + \frac{2}{2 K_\alpha \SNR}
\end{equation*}
where
\begin{equation*}
   l_\circ^*(r,K_\alpha;\SNR)
   = \frac{1 + \sqrt{1 + 4 K_\alpha^2 r^2 \SNR^2}}{2 K_\alpha \SNR}
\end{equation*}
which is independent of $d$.  Hence, taking the partial derivative of
$ E_\mathrm{bnd}^\lambda(r,K_\alpha,l,d,\beta^*_\lambda;\SNR)$ with
respect to $d$ and substituting $l$ with $l_\circ^*(r,K_\alpha;\SNR)$
one can show that this partial is zero if
\begin{equation*}
  d_\circ^*(r) = \sqrt{2} r.
\end{equation*}
Hence, if $\beta_\lambda^*= \beta_\circ^*(r,d,l ; \SNR )$ then one can show
\begin{equation*}
   l_\lambda^*(r,\alpha;\SNR) =  l_\circ^*(r,K_\alpha;\SNR) \,
   \textnormal{and } \,
   d_\lambda^*(r,l,\alpha;\SNR) =  d_\circ^*(r).
\end{equation*}

The case where $\beta_\lambda^* = {d^2}/{4 l^2} \cdot \left( l -
K_\alpha \right)$ is a bit more tedious.  In the sequel we provide a
general derivation for the $l$ and $d$ that maximize the union bound
for a code that has been expurgated so that the minimum distance is at
least $d_{\Omega}(R)$.  In this direction we let
\begin{equation*}
    \Et_\mathrm{bnd}^\lambda(r,K_\alpha,l,d;\SNR)
    =  E_\mathrm{bnd}^\lambda\left(r,K_\alpha,l,d,\frac{d^2}{4 l^2}(l-K_\alpha);\SNR\right).
\end{equation*}
It is simple to check via direct substitution that
\begin{align}
    &\Et_\mathrm{bnd}^\lambda(r,K_\alpha,l,d;\SNR)  \notag \\
    &\quad = \frac{d^2}{8 l^2} (K_\alpha-l)^2 \cdot \SNR -
       \log\left[
           \frac{d}{1+K_\alpha} \sqrt{ 1 - \frac{d^2}{4 l^2} }
           \right] - R.  \label{eqn:union_wK}
\end{align}
Examining \eqref{eqn:union_wK} it is easy to see that if the $l$
and $d$ that maximize the union bound are equal to $1+K_\alpha$ and
$d_{\Omega}(R) (1+K_\alpha)$ respectively then \eqref{eqn:union_wK} is
exactly the union bound for a spherical code with minimum distance
$d_{\Omega}(R)$ in the expurgated regime.   We now show that there
exists a scaling $\alpha$ such that this is true.  That is, we solve
for the $\alpha$ such that the $l$ and $d$ that maximize the union
bound are equal to $1+K_\alpha$ and $d_{\Omega}(R) (1+K_\alpha)$
respectively.

To begin we introduce a Lagrange multiplier $\mu$ and consider minimizing
\begin{equation*}
    \Et_\mathrm{bnd}^\lambda(r,K_\alpha,l,d;\SNR)
    + \mu \left( d_{\Omega}(R) - d/(1+K_\alpha) \right).
\end{equation*}
It can be shown that in the regime of interest $\mu > 0$,
$d_\lambda^* = d_{\Omega}(R) (1+K_\alpha)$ and
$l_\lambda^*(r,\alpha;\SNR)$  satisfies
\begin{equation*}
   \frac{d_{\Omega}(R)^2}{4} = \frac{(l_\lambda^*)^2
   +\left(K_\alpha^2 \cdot (l_\lambda^*)^2-K_\alpha \cdot (l_\lambda^*)^3\right) \, \SNR }{
   K_\alpha \, (1+K_\alpha)^2 \, (K_\alpha-l_\lambda^*) \, \SNR .
                        }.
\end{equation*}
In order to minimize the probability of error we are left to maximize
\begin{align}
   &2 \Et_\mathrm{bnd}^\lambda(r,K_\alpha,l_\circ^*(r,K_\alpha;\SNR),d_{\Omega}(R) (1+K_\alpha);\SNR) \notag \\
   &\quad = 1 - 2 R + K_\alpha^2 \SNR
  + \frac{l_\lambda^*}{K_\alpha \left((l_\lambda^* K_\alpha - 2 K_\alpha^2)\SNR -1\right) } \notag \\
   &\qquad{} - \log\left(
  \frac{4 (l_\lambda^*)^2 \left((l_\lambda^* K_\alpha - K_\alpha^2)\SNR -1\right)
       }{K_\alpha^2 (1+K_\alpha)^2 \left(l_\lambda^*-K_\alpha\right)^2}
  \right)  \notag
\end{align}
as a function of $K_\alpha$.  It is straightforward\footnote{Recall that
$l_\lambda^*$ is a function of $K$} to check that
$\Et_\mathrm{bnd}^\lambda(r,K_\alpha,l_\circ^*(r,K_\alpha;\SNR),d_{\Omega}(R) (1+K_\alpha);\SNR)$ is maximized when
\begin{equation*}
   K_\alpha = K^*_\alpha(d_{\Omega}(R);\SNR)
            = \frac{4}{\left(4-d_{\Omega}(R)^2\right) \, \SNR},
\end{equation*}
yielding
\begin{align*}
&l_\lambda^*(r,1/(1+K^*_\alpha(d_{\Omega}(R);\SNR));\SNR) \\
&\qquad = 1 + K^*_\alpha(d_{\Omega}(R);\SNR));\SNR).
\end{align*}
Substituting this into \eqref{eqn:Ecirc_def} we have
\begin{align}
   &\Et_\mathrm{bnd}^\lambda(r,K^*_\alpha,l_\lambda^*,d_{\Omega}(R) (1+K^*_\alpha);\SNR) \notag \\
   &\quad = \frac{\SNR}{8} \cdot d_{\Omega}(R)^2 - \log\left( d_{\Omega}(R) \sqrt{1 - d_{\Omega}(R)^2/4}
                                          \right) - R \notag
\end{align}
which is precisely the error exponent for the union bound using a
spherical code with minimum distance $d_{\Omega}(R)$ in the expurgated
regime.

\section{Derivation of \eqref{eqn:prop_2} with a Spherical Region}
      \label{sec:bound}
      \label{sec:sphere_details}

We have now provided a geometric characterization of the typical error
events for the AWGN error exponent.   For this derivation we used cone
$\Rcone(\theta_\mathrm{AWGN})$ for the region $\cR$ in
\eqref{eqn:gallager_region}.   Recall that in order to show that the
mod-$\Lambda$ channel can achieve the sphere-packing error exponent
for the AWGN channel we took a scaling $\alpha$ that corresponded to a
valid sphere.   We extend our previous definition of ``valid'' regions
[cf~\eqref{eqn:valid_sphere_def}] to be the regions $\cR$ such
that:
\begin{align}
&1.  \quad \pr\left(\bc + \bz \notin \cR \right) \expeq
\pr\left(\bc + \bz \notin \Rcone(\theta_\mathrm{AWGN}) \right) \\
&2.  \quad \cR \subset \Rcone(\theta_\mathrm{AWGN})
\end{align}
We now show that by using the valid sphere
$\Rsphere(\theta_\mathrm{AWGN})$ one may obtain the best known lower
bounds on the
error exponent.  We then use this derivation to show that the mod-$\Lambda$
channel also can achieve the random coding error exponent
$E^r_\mathrm{AWGN}(R)$.

\begin{figure*}
\begin{center}
   \begin{tabular}{cc}
      \begin{psfrags}
      \psfrag{A}[cc]{$\frac{1}{\alpha}$}
      \psfrag{M}[lb]{$\be_y$}
      \psfrag{ME}[lt]{$\bc_e$}
      \psfrag{D}[lc]{$\cD_s(d,\alpha)$}
      \psfrag{D2}[rb]{$\frac{d}{2}$}
      \psfrag{TA}[lc]{$\theta$}
      \psfrag{TD}[bc]{$\thetaD{d}$}
      \psfrag{X}[cc]{$x$}
      \psfrag{Y}[cc]{$y$}
      \psfrag{G}[cc]{}
      \psfrag{B}[cc]{$\beta_s$}
      \psfrag{OA}[tc]{$1-\alpha$}
      \psfrag{ER}[lc]{$e^{-R}$}
      \includegraphics[scale=0.592]{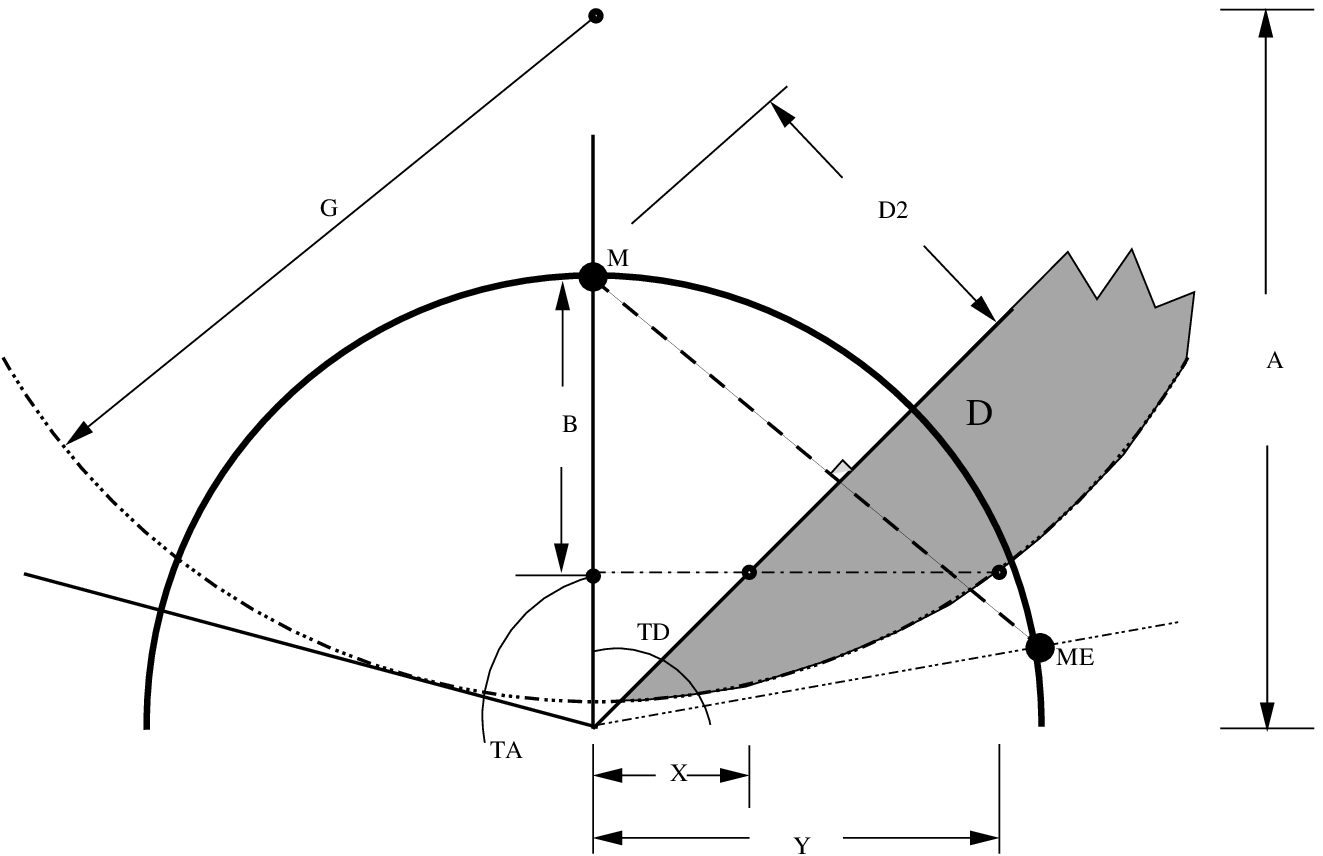}
      \end{psfrags}
&
      \begin{psfrags}
      \psfrag{G}[rc]{}
      \psfrag{A}[lc]{\small $\frac{1}{\alpha}$}
      \psfrag{D}[lc]{$\cD_s(d,\alpha,\beta)$}
      \includegraphics[scale=1.72]{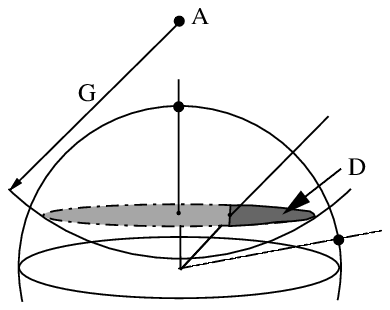}
      \end{psfrags}
   \end{tabular}
\end{center}
   \caption{The parameters for the derivation of the AWGN error exponent using
     spherical regions.   (a)
     A 2D representation of the bounding technique.   The region
     $\cD_s(d)$ corresponding
     to an error with a codeword at distance $d$ condition on the event that
     the noise remains in $\cR$ can be seen shaded in gray.  (b)
     A three dimensional representation of the  region corresponding
     to an error with a codeword at distance $d$ condition on the event that
     the noise remains in $\cR$ and the radial component of
     the noise.}
   \label{fig:big_picture_bound}
\end{figure*}

Again using the tangential sphere bound we consider the intersection
of the region $\Rsphere(\theta)$ with the half space that
orthogonally bisects a cord of length $d$ that has one end point at
transmitted codeword and the other at $\bc_e$.  This is the shaded
region in Fig.~\ref{fig:big_picture_bound}.   We let
$\cD_s(d,\alpha)$ be this intersection.   We also let
$\cD_s(d,\theta,\beta_s)$ be the intersection of
$\cD_s(d,\alpha)$ with the hyperplane, say $\cH$,
such that $\be_y' \bx = \beta_s \sqrt{n P}$ for all $\bx
\in \cH$.   The $n-1$ dimensional region
$\cD_s(d,\alpha,\beta_s)$ can be seen in
Fig.~\ref{fig:big_picture_bound}.   We may bound
$\Prmub^r$ as in \eqref{eqn:cone_bound} yielding
\begin{align}
    &\Prmub^r
    \exple  \max_{ 0 \le d \le 2 }
    \pr\left( \sqrt{nP} \cdot \be_y + \bz \in \cD_s(d,\alpha)
     \right)  \notag \\
    & \qquad \qquad{} \times \exp{\left(n R + (n-1) \log
    \left( d \sqrt{1 - \frac{d^2}{4} } \right) \right) }
   \label{eqn:sphere_bound}
\end{align}

In order to obtain an exponential bound to $\pr\left( \sqrt{nP} \cdot
\be_y + \bz \in \cD_s(d,\alpha) \right)$ we again integrate along the
radial component of the noise.   This can be seen in
Fig.~\ref{fig:big_picture_bound}.  When the radial component of the
noise has magnitude $\beta_s$, the probability $\pr\left( \sqrt{nP}
\cdot \be_y + \bz \in \cD_s(d,\alpha) \right)$ is simply the
probability that the second component is greater than $x_s$ while the
magnitude of the second thorough $n$th component is less than $y_s$,
where
\begin{equation}
   x_s = x_s(\beta_s, \thetaD{d}) = x_c(\beta_s, \thetaD{d})
\end{equation}
and
\begin{equation}
 y_s^2 = y_s^2(\alpha,\beta_s) = \sqrt{n P} \left(
            \frac{\sin^2 \theta}{\alpha^2}
            - \left( \frac{1}{\alpha} - (1 + \beta_s)\right)^2
        \right)
\label{eqn:dist_bound}
\end{equation}
and can be derived through the
the geometry in Fig.~\ref{fig:big_picture_bound}.

Thus, applying Proposition~\ref{prop:chernoff_constrain}
we have, similar to \eqref{eqn:cone_exponent},
\begin{align}
&\pr\left( \sqrt{nP} \cdot \be_y + \bz \in
 \cD_s(d,\alpha) \right) \notag \\
 &\quad \le \max_{  \beta_s \ge -1 }
          \exp\bigl( -n \,
          {\Ebd}\left(\beta_s,x_s,y_s; \SNR \right) \bigr)
 \label{eqn:sphere_exponent}
\end{align}
where $\Ebd$ was defined in \eqref{eqn:expon_Ebd}.

It is clear from the geometry in Fig.~\ref{fig:big_picture_mlan} and
Fig.~\ref{fig:big_picture_bound} that the error exponent using
spherical regions is exactly that of the mod-$\Lambda$ channel if $l =
1 + K_\alpha$.   Hence, if $\alpha = \alpha_\lambda^*$ the geometry of
the error events coincide.   That is, $\beta_s^* = \beta_\lambda^*$ and
$d_s^* = d_\lambda^*$.  Clearly this analysis extends to the analysis
using a cone so that typical error events for the AWGN channel using
$\Rcone(\theta_\mathrm{AWGN})$ and the typical events using
$\Rsphere(\theta_\mathrm{AWGN})$ {coincide}.

\section{Derivation of \eqref{eqn:cone_bound}}
   \label{append:dist_dist}

For a fixed codebook
$\cC$ and fixed codeword $\bc \in \cC$ we have from
\eqref{eqn:gallager_region}
\begin{align}
  \Prmu^r(\bc) &=
 \sum_{ \bc_\mathrm{e} \ne \bc } \pr\left(
                         \| \by - \bc_e \| \leq
                         \| \by - \bc \|
                         , \bc + \bz \in \cR
                      \right) \notag \\
 &= \sum_{ \bc_\mathrm{e} \ne \bc } \pr\left(
                       \bc + \bz \in
                       \cD_c\left(\| \bc - \bc_e \| \right)
                       \right).  \label{eqn:need_dist}
\end{align}
Note that \eqref{eqn:need_dist} is a function of the \emph{distance}
between the transmitted codeword and all other codes words.   Recall
that for a fixed code $\cC$ the distance distribution or
spectrum of the code is defined as follows.   Let $b_{\bc}(s,t)$ be the
number of codewords of $\cC$ that are at least a distance $s$
from $\bc$ but not further than $t$.   That is,
\begin{equation*}
   b_{\bc}(s,t) \myDef | \{ \bc_e \in \cC \, : \,
                              s \le \| \bc - \bc_e \| < t \} |
\end{equation*}
We further let $b(s,t)$ be the average of $ b_{\bc}(s,t)$ over the code.
That is,
\begin{equation*}
   b(s,t)  \myDef \frac{1}{|\cC|} \sum_{\bc \in \cC}
                  b_{\bc}(s,t).
\end{equation*}
For a spherical code $0 < \| \bc - \bc_e \| < 2 \sqrt{nP}$.
Thus, by discretizing the interval $[0,2 \sqrt{nP} ]$ into intervals of length
$\Delta$,
we may upper bound \eqref{eqn:need_dist} as
\begin{equation*}
    \Prmu^r(\bc) \le
                       \sum_{i=0}^{k_\Delta}
                       b_{\bc}(i\Delta, (i+1) \Delta) \pr\left(
                       \bc + \bz \in
                       \cD_c\left( i \Delta \right)
                       \cup
                       \cD_c\left( (i+1) \Delta \right)
                       \right)
\end{equation*}
where $k_\Delta = {\lceil 2 \sqrt{nP}/\Delta \rceil - 1}$.   By
spherical symmetry and linearity of expectation we have
\begin{align}
  \Prmub^r &\le \E  \sum_{i=0}^{k_{\Delta}}
                       b( i\Delta, (i+1) \Delta) \pr\left( \sqrt{nP} \cdot
                        \be_y + \bz \in
                       \cD_c\left(i \Delta \right)
  \cup
                       \cD_c\left( (i+1) \Delta \right)
                       \right) \\
       &=   \sum_{i=0}^{k_\Delta}
                       \pr\left(  \sqrt{nP} \cdot
                        \be_y + \bz \in
                       \cD_c\left(i \Delta \right)
  \cup
                       \cD_c\left( (i+1) \Delta \right)
                       \right)
                       \E \,
                       b( i\Delta, (i+1) \Delta).
\end{align}
In the limit $\Delta \to 0$ we have that $ \E \, b(i\Delta, (i+1) \Delta)$ is
proportional to the radius of the spherical cross section at a height
$d = i \Delta$, i.e.,
\begin{equation*}
\E \, b(i\Delta, (i+1) \Delta) \to
   \sin  \thetaD{d} = d \sqrt{1 - \frac{d^2}{4}}
\end{equation*}
since the codewords are chosen uniformly over the surface of the sphere.
Thus,
\begin{align}
    \Prmub^r
    &\le K
            \int_{0}^2 e^{n R} \left( s \sqrt{1 - \frac{s^2}{4} } \right)^{n-1}
   \pr\left( \be_y + \bz \in \cD_c(s)
  \cup
                       \cD_c\left( (i+1) \Delta \right)
\right)\,ds  \notag \\
    &\le 2 K  \max_{ 0 \le d \le 2 } e^{n R}
    \left( d \sqrt{1 - \frac{d^2}{4} } \right)^{n-1}
   \pr\left( \be_y + \bz \in \cD_c(d) \right)
\end{align}
where $K$ is a normalizing constant.

\section{Derivation of \eqref{eqn:mlan_bound}}
   \label{append:dist_dist_lattice}

Recall that in order to derive bounds for $\Prmub^r$ we used the spectrum of the ensemble of codes.   Now, we
use a random coding argument to derive the error probability for the
mod-$\Lambda$ channel.   We consider the ensemble of random codes that
are $\it{i.i.d}$ and uniform over the Voronoi of the lattice.   Recall
that in order to derive bounds for $\Prmub^r$ we
used the spectrum of the ensemble of codes.  In this direction, let
$b_{\bc}(s,t,\theta_{d_1},\theta_{d_2})$ be the number of codewords of
$\cC$ that are at least a distance $s$ from the transmitted
codeword $\bc$ but not further than $t$ and form an angle with the
dither between $\theta_{d_1}$ and $\theta_{d_2}$.   That is,
\begin{equation*}
   b_{\bc}(s,t,\theta_{d_1},\theta_{d_2})
       \myDef \left| \left\{ \bc_e \in \cC \, : \,
                              s \le \| \bc - \bc_e \| < t
          \text{ and } \cos \theta_{d_2} \le
          \frac{\bu \bc_e^\dagger}{\|\bu\| \|\bc_e\|} \le \cos \theta_{d_1}
          \right\} \right|
\end{equation*}
We further let $b(s,t,\theta_{d_1},\theta_{d_2})$ be the average of
$ b_{\bc}(s,t,\theta_{d_1},\theta_{d_2})$ over the code.
By appropriately discretizing by $\Delta$ and taking the limit it can be shown
that
\begin{equation*}
\lim_{\Delta \to 0}  b_{\bc}(d,d+\Delta,\theta_{p},\theta_{p} + \Delta ) =
    \alpha e^{nR} \left( {d} \sin \theta_p \right)^{n-1}
\end{equation*}
where the $\alpha$ appears due to the scaling of the lattice.
Thus, \eqref{eqn:union_lattice} becomes
\begin{equation}
   \Prmub^{\lambda,r}
    \exple
   \mathop{ \max_{ (d,l) } }_{  0 \le d \le 2 r, l \ge K_\alpha }
    e^{n R}
    \left( \alpha d \sqrt{1-\frac{d^2}{4 l^2}} \right)^{n-1}
   \pr\left( \frac\bv{\alpha} + \bz_\mathrm{eff}'
            \in \cD_\lambda(r,d,l) \right)
\end{equation}

\section{Proof of Lemma \ref{lem:z}}
\label{sec:z_proof}

It is simple to check that
\begin{equation*}
z(K;d,R,\SNR) = 2 K \cdot ( c(d;R,\SNR) - E_\mathrm{sp}(R(\theta_\zeta(K;\SNR));\SNR) )
\end{equation*}
where $c(d;R,\SNR) \ge 0$ if $d \ge 0$, $R \ge 0$ and $\SNR > 0$, and
is independent of $K$.  Furthermore, $2 K \cdot
E_\mathrm{sp}(R(\theta_\zeta(K;\SNR));\SNR) \ge 0$ and is
monotonically increasing on the interval $[1/\SNR,\infty)$ as a
function of $K$.  Hence, since
$E_\mathrm{sp}(R(\theta_\zeta(1/\SNR;\SNR));\SNR) =0$ the equation
$E_\mathrm{sp}(R(\theta_\zeta(K;\SNR));\SNR) = c(d;R,\SNR)$ has a
unique solution for $K$ and thus $z(K;d,R,\SNR)$ has one root on the
interval $[1/\SNR,\infty)$.

\bibliographystyle{IEEEtran}
\bibliography{swannackALL}

\end{document}